\documentclass[twocolumn,floatfix]{aastex62}
\pdfoutput=1
\usepackage{amsmath,amstext,multirow}
\usepackage[T1]{fontenc}
\usepackage{apjfonts,xcolor}
\usepackage[figure,figure*]{hypcap}

\def\be{\begin{eqnarray}}   \def\ee{\end{eqnarray}}
\def\ben{\begin{equation}\begin{aligned}} \def\een{\end{aligned}\end{equation}}
\shorttitle{The Age Metallicity Relation}
\shortauthors{Sharma et al.}
\newcommand{\feh}{${\rm [Fe/H]}$}
\newcommand{\alphafe}{$[\alpha/{\rm Fe}]$}
\newcommand{\fehalpha}{$([{\rm Fe/H}], [\alpha/{\rm Fe}])$}
\begin{document}
\title{
Chemical Enrichment and Radial Migration in the Galactic Disk - the origin of the [$\alpha/\rm Fe]$ Double Sequence.
}

\author{Sanjib Sharma}
\affiliation{Sydney Institute for Astronomy, School of Physics, The University of Sydney, NSW 2006, Australia}
\affiliation{ARC Centre of Excellence for All Sky Astrophysics in Three Dimensions (ASTRO-3D)}
\author{Michael R. Hayden}
\affiliation{Sydney Institute for Astronomy, School of Physics, The University of Sydney, NSW 2006, Australia}
\affiliation{ARC Centre of Excellence for All Sky Astrophysics in Three Dimensions (ASTRO-3D)}
\author{Joss Bland-Hawthorn}
\affiliation{Sydney Institute for Astronomy, School of Physics, The University of Sydney, NSW 2006, Australia}
\affiliation{ARC Centre of Excellence for All Sky Astrophysics in Three Dimensions (ASTRO-3D)}
\begin{abstract}
The $([\alpha/{\rm Fe}],[{\rm Fe/H}])$ distribution of Milky Way stars shows at least two distinct sequences, which have traditionally been associated with the thin and thick disc components. The abundance distribution  varies systematically with location $R$ and $|z|$ across the Galaxy. Using an analytical chemodynamical model that includes the effects of radial migration and kinematic heating, we show that it is possible to reproduce the observed abundance distribution at different locations.  Unlike some earlier models, our scheme does not require a distinct thick disc component emerging from a separate evolutionary path. The proposed model has a continuous star formation history and a continuous age velocity dispersion relation. Moreover, [$\alpha$/Fe] is constant for stellar ages less than 8 Gyr, but increases sharply for older stars over a time scale of 1.5 Gyr. The gap between the two sequences is due to this sharp transition. We show that the high-\alphafe{} sequence at the low metallicity end is simply a pile-up of old stars. At the high metallicity end, we find a sequence of stars having different ages with a similar birth radius that originated in the inner disc. Our model successfully explains the uniformity of the locus of the high-\alphafe{} sequence across different locations.
The low-\alphafe{} sequence contains stars with different birth radii that owes its existence to radial migration. For the low-\alphafe{} sequence, angular momentum is anti-correlated with [Fe/H], and the model can reproduce this trend at different Galactic locations.
If radial migration is not included, the model fails to generate the double sequence and instead shows only a single sequence. Our simple scheme has major advantages over earlier chemodynamical models, as we show.

\end{abstract}

\section{Introduction}
The formation and evolution of the Milky Way disk is one of the outstanding questions facing Galactic archaeology today \citep{2002ARA&A..40..487F}. However, new observational data are providing avenues to resolve this longstanding question. With the advent of the \textit{Gaia} satellite \citep{2016A&A...595A...1G,2018A&A...616A...1G} and large-scale spectroscopic surveys such as
LAMOST \citep{2012RAA....12..723Z},
RAVE \citep{2020arXiv200204377S}, Gaia-ESO \citep{2012Msngr.147...25G}, APOGEE \citep{2017AJ....154...94M}, and GALAH \citep{2015MNRAS.449.2604D}, our understanding of the Galaxy is in the midst of a revolution. Astrometric parameters from \textit{Gaia} \citep{2018A&A...616A...1G,2018A&A...616A...2L} allow improved phase space estimates for more than a billion stars, while large-scale spectroscopic surveys allow reliable chemical abundance determinations and ages, giving us an unprecedented picture of the Galaxy and the ability to trace its structure and evolution through time.

Ever since the discovery of the thick disc by \citep{1983MNRAS.202.1025G}, its origin and its link to
the thin disc, which makes up most of the stars in the Milky Way, has remained unclear. Further studies of the thick disc
stars have revealed that they are different from the thin disc star in multiple ways, which lead to the notion
that it might be distinct from the thin disc and might
have originated from a separate evolutionary pathway.
The thick disc was initially identified in observations of star counts in the solar neighbourhood away from the mid-plane of the disk. A single exponential could not fit the observed stellar distributions; two components were required: a thin disk component with a small scale height ($\approx$ 300 pc), and a thick disk component with a three-fold increase in scale height \citep{1983MNRAS.202.1025G}.
Spectroscopic observations of stars high above the plane belonging to the thick disk reveal that these stars are older and have higher \alphafe{} relative to stars in the plane (e.g., \citealt{1998A&A...338..161F,2007ApJ...663L..13B,2013A&A...560A.109H}). Hence, the thick disk has increasingly been identified via stellar chemistry, rather than a star's distance from the plane. Large spectroscopic surveys
have enabled a detailed exploration of the distribution of stars in the \fehalpha{} plane and at different $R$ and $|z|$ locations across the Galaxy \citep[][also available in the text later]{2015ApJ...808..132H}.
In the solar annulus, the distribution shows two major sequences, a high-\alphafe{} sequence (associated with the thick disc) and a low-\alphafe{} sequence (associated with the thin disc). The sequences are almost parallel at the low \feh{} end, but as \feh{} increases the two sequences progressively come closer and merge at slightly super-solar metallicities. It is not clear as to where the high-\alphafe{} track ends. The higher \feh{} stars of the high-\alphafe{} track have kinematics similar to that of the thin disk
\citep{2007ApJ...663L..13B,2011A&A...535L..11A,2017A&A...608L...1H,2020arXiv200303316C}.
The distribution of stars in the low-\alphafe{} sequence changes systematically with location $R$ and $|z|$ across the Galaxy. However, the locus of the high-\alphafe{} sequence appears to be the same at all locations to within observational errors \citep{2014ApJ...796...38N,2015ApJ...808..132H}.
Several observation studies have found that the high-\alphafe{} stars identified with the thick disk have a short scale-length, and only extend out to roughly the Sun's position. At larger radii, these high-\alphafe{} populations are absent and stars above the plane are instead made up of flaring solar-\alphafe{} populations  (e.g., \citealt{2011ApJ...735L..46B,2012ApJ...753..148B,2015ApJ...808..132H,2019ApJ...874..102W}). For a detailed discussion of flaring see \citet{2015ApJ...804L...9M} and
for a schematic illustration of the thin and thick discs see Fig. 1 of \citet{2019MNRAS.486.1167B}.
In the inner disc, we find only a single sequence, the low-\alphafe{} sequence has shifted towards higher \feh{} and is merged with the high-\alphafe{} sequence, implying that the thin and thick disc are chemically connected \citep{2014ApJ...781L..31S,2015ApJ...808..132H,2016A&A...589A..66H}. The distinct gap between the chemical thin and thick disks found locally does not exist in the inner Galaxy, leaving its origin in the solar neighbourhood an open question. The thick disc has also been associated
with distinct kinematic features. The velocity dispersion as a function of age for stars in the solar neighborhood shows a break from a power law with an increase being found for older stars
\citep{1991dodg.conf...15F, 1993A&A...275..101E, 2001ASPC..230...87Q,2020arXiv200406556S}.

Despite the thick disc appearing to be a distinct population, a number of studies have argued against it. \citet{1987ApJ...314L..39N} suggested that the
double exponential vertical distribution of stars can be explained by a disc whose vertical velocity dispersion (or equivalently scale height) varies continuously with metallicity. \citet{2012ApJ...751..131B} binned up the SEGUE \citep{2009AJ....137.4377Y} G-dwarfs in the \fehalpha{} plane to create mono-abundance populations (MAP) and studied the spatial distribution of stars for each of the MAPs. Using these MAPs they showed that the mass weighted distribution of scale height in the solar annulus is continuous. A similar analysis was repeated using
giants from APOGEE by \citet{2016ApJ...823...30B}
and \citet{2017MNRAS.471.3057M} that lead to the same conclusion.
\citet{2020arXiv200406556S} argue that the velocity dispersion of the thick disc stars follow the same relations for their dependence on age, angular momentum and metallicity, as stars belonging to the thin disk. The apparent uniqueness of the thick disc kinematics is because the velocity dispersion also depends on angular momentum, this was not taken into account in the previous studies.

\citet{2009MNRAS.396..203S} showed using a chemical
evolution model that the double sequence (bimodality) in the \fehalpha{} plane can be reproduced by a model having a continuous star formation history.
They introduced a chemical
evolution model that included radial flow of gas
and radial migration of stars in addition to
other relevant physical processes, e.g., star formation history, stellar yields, gas accretion and outflow.
In addition to chemical abundances
their model also included
the phase space distribution of stars.
The free parameters of the model were tuned to reproduce observations of the solar neighbourhood from the Geneva-Copenhagen Survey \citep{2004A&A...418..989N}.
Their model was able to reproduce many observed properties of the disk, anti-correlation
of angular momentum with metallicity for the low-\alphafe{} stars \citet{2009MNRAS.399.1145S}, the scatter in the age-metallicity relation locally, as well as the shift in both the peak of the MDF (radial gradient) and change in  shape  with radius. The origin of the double sequence was attributed to the
time delay of SNIa, which makes the \alphafe{} transition from high-\alphafe{} at earlier times to low-\alphafe{} at later times. Radial migration  was
identified as the key mechanism that brought kinematically hot stars from the inner disc to the solar neighborhood to create the thick disc. The large spread in metallicity of the low-\alphafe{} sequence (thin disc) was also due to radial migration coupled with the existence of a strong metallicity gradient in the ISM.
Later findings
\citep{2014ApJ...794..173V,2018MNRAS.476.1561D} that radial migration is efficient only for kinemtaically cold stars,
raised questions as to how can the
thick disc be created by migration of stars from the inner disc and still be  kinematically hot.
\citet{2017MNRAS.470.3685A} show using idealized N-body simulations that if the Galaxy has an inside out growth, it
is possible to have outward migrators with high velocity dispersion.

In spite of significant successes of the
\citet{2009MNRAS.399.1145S} model, their
study lacked a detailed comparison with observations. They only made predictions
for the \fehalpha{} distribution in the solar neighborhood and compared it with the limited observational data set that was available at that time.
Although two sequences could be clearly seen but their model also predicted a significant number of stars in between the two sequences. Unfortunately, a large enough kinematically unbiased
data set was not available to thoroughly
test the predicted distribution of stars in the \fehalpha{} plane.
With data from APOGEE survey becoming available, it soon became possible to study the distribution of \fehalpha{}
at different $R$ and $|z|$ locations of the Galaxy \citep{2014ApJ...796...38N, 2015ApJ...808..132H}. Additionally, the bimodality was clearly visible in this kinematically unbiased data set. The \fehalpha{} distributions showed a number of interesting trends with Galactic location
which were never compared with model predictions; this comparison is one of the main aims of this paper.

Since \citet{2009MNRAS.396..203S}, a number of other chemical evolution models have been proposed, but no attempt has been made
to reproduce the observed \fehalpha{} distribution at different Galactic locations.
Models by \citet{2013A&A...558A...9M,2014A&A...572A..92M} use cosmological zoom-in simulations as input to generate realistic kinematic distributions, and then add a detailed chemical evolution prescription on top of the dynamics from the simulation. However, the \fehalpha{} distribution does not show bimodality, and the \feh{} distribution peaks at
the same value independent of the radius $R$.
Such simulations are computationally expensive to run and often difficult to directly compare to the Milky Way, as the simulated galaxy might not be a perfect match for the evolutionary history of the Milky Way. A potential way around this is to characterize the dynamic processes found in N-body or cosmological simulations with analytic functions, allowing for models that have good approximations for the important physical processes in the dynamical evolution of a disk, while also being inexpensive and having the flexibility to tune parameters to better match Milky Way observables, e.g., as in \citet{2015A&A...580A.126K,2015A&A...580A.127K}.
In this model, the old stars forms the high-\alphafe{} sequence and the
young stars form the low-\alphafe{} sequence similar to the observed sequences. However, a proper distribution of stars in \fehalpha{} plane with two distinct sequences was not shown.

There have also been alternate chemical evolution
models that suggest a different formation scenario for the thick disc. A typical feature of these is strong star formation at early times which forms the thick disc, followed by a period where  star formation is quenched or stops, the star formation resumes and continues at a slower rate at later times.
Some models have closed box chemical evolution
\citealt{2015A&A...578A..87S, 2016A&A...589A..66H,2019A&A...625A.105H} while others have open box chemical evolution where accretion of fresh gas happens over an extended time scale \citep{1997ApJ...477..765C,2001ApJ...554.1044C, 2019A&A...623A..60S}.
A common problem with these models is that they either ignore
radial migration or consider it to be insignificant \citep{2019A&A...625A.105H}. Typically, the chemical evolution tracks for a given birth radius show discontinuity or abrupt changes both in the the \fehalpha{} plane and evolution with time. \citet{2019A&A...623A..60S}, which has a longer delay between the first and the second gas infall phase, leads to a loop in \fehalpha{} plane, which does not match observations. While the evolutionary tracks have been shown to qualitatively coincide with the locus of the high and low-\alphafe{} sequences, a detailed prediction of the
distribution of stars in the \fehalpha{} plane and its variation with $R$ and $|z|$ has not been done. Additionally, the anti-correlation of angular momentum with metallicity has
also not been shown.

There are also some general purpose chemical evolution models
which track the evolution in individual radial zones and have up-to-date stellar yields and realistic time delays for SNIa  \citep{2014ApJ...796...38N, 2017ApJ...835..224A}
but they lack phase-space information. They are potentially useful diagnostic tools in determining the impact of star formation efficiencies and gas infall and outflow rates on the chemical history of the Galaxy.

Full blown chemical evolution models, especially those that include dynamical processes like \citet{2009MNRAS.396..203S} model, are not easy to
fit to data. In this regard, models based on an analytical distribution function offer a distinct advantage.
Phase space distribution functions based on
actions have been developed \citet{2012MNRAS.426.1328B}.
\citet{2015MNRAS.449.3479S} extended analytical action based distribution function to also
track the evolution of metallicity; additionally,
they introduced a prescription for radial migration.
However, they did not track the evolution of any element other that iron.

None of the previous models have been shown to reproduce the observed trends Galaxy wide in their entirety. There are various reasons for why this is the case, varying from outdated observational constraints on the chemical distributions of the Galaxy to overly simplistic approximations for the velocity dispersion of the disk. However, significant improvements have been made in the characterization of the velocity dispersion and its dependence on age, angular momentum, and \feh{} in \citet{2020arXiv200406556S}. This, along with the improved observational constraints provided by \textit{Gaia} and large-scale spectroscopic surveys, allows us to generate a new chemodynamical model for the Galaxy and make detailed
comparisons of it to observations.

In this paper, we will describe the framework of our chemodynamical model which take into account a number of   relevant physical processes and reproduces chemodynamical observations throughout the disk. This paper is organized as follows: in Section 2, we describe the observational data sets used to constrain our model. In Section 3, we describe the parameters and functionary of the model.
In Section 4, we describe our results and directly compare them with observational data sets. In Section 5, we discuss impact of our work and its ability to reproduce observational trends throughout the Galaxy, as well as compare to existing works on the chemodynamical evolution of the disk.
Section 6 summarizes our main findings and highlights where future improvements can be made.

\section{Data}
For studying the distribution of stars in the \fehalpha{} plane at different $R$ and $|z|$ locations, we use data from  the APOGEE survey. We use the ASPCAP catalog of stellar parameters and abundances from APOGEE-DR14.
We select
stars according to the following criteria,
\be
(1.0<\log g<3.5)\&(3500 <T_{\rm eff}/{\rm K}<5300)\& \nonumber\\
(7<H<11).
\label{equ:rgselect1}
\ee
The $\log g$ and $T_{\rm eff}$ selection function is designed to select giants. Although the APOGEE survey extends in $H$ band upto  13.8 mag, but we restrict to 11 mag as  beyond it the selection function is not homogeneous and difficult to reproduce.
Additionally, we restrict to stars with
${\rm S/N} > 80$, are main survey
targets (EXTRAARG flag = 0), are not flagged bad (ASPCAPFLAG $\neq 23$) and have valid distance, [Fe/H] and \alphafe{}. This resulted in a sample of 94,488 stars.  For distances, we use the BPG distances  by \citet{2016A&A...585A..42S} from the APOGEE-DR14 value added catalog.
We calibrate our chemeical enrichment model using data from
LAMOST and GALAH. Since, the abundance estimates from different spectrosocpic surveys in general do not agree with each other we need to
calibrate them. We have crudely recalibrated the APOGEE \feh{} values to
that of GALAH and LAMOST by decreasing them by -0.15 dex.

For studying the dependence of angular momentum on metallicity, we use
red giant stars from both the APOGGE and LAMOST surveys.
We used the LAMOST-DR4 value added catalog from
\citet{2017MNRAS.467.1890X}, for  radial velocity, $T_{\rm eff}$, $\log g$, [Fe/H], [$\alpha$/Fe], and distance.
For the RG stars, we adopt the following selection criteria,
\be
(1<\log g < 3.5) \& (3500\; <\; T_{\rm eff}/{\rm K}\;<5500)\&\nonumber\\
(7<H<13.8).
\label{equ:rgselect2}
\ee
The criteria is less strict than \autoref{equ:rgselect1}, so as to increase the sample size.

To model the chemical enrichment we make use of data from
LAMOST and GALAH surveys. We used the LAMOST-DR4 value added catalog from
\citet{2017MNRAS.467.1890X}.
For LAMOST stars, we used two types of stars, the MSTO stars and the red-giant (RG) stars. The ages for the
LAMOST-MSTO sample were taken from \citet{2017ApJS..232....2X}
and for the LAMOST-RG-CN sample were taken from \citet{2019MNRAS.484.5315W}.
The LAMOST-RG-CN
sample consists of red giant branch stars with ages derived from
spectroscopic C and N features.
For the GALAH survey, we used MSTO stars.
More precisely, we make use of the extended GALAH catalog (GALAH+), which also includes data from TESS-HERMES \citep{2018MNRAS.473.2004S} and K2-HERMES \citep{2019MNRAS.490.5335S} surveys
that use the same spectrograph and observational setup as the GALAH survey.
In this paper, we exploit parameters from GALAH-iDR3, an internal data release where every star has been analysed using SME and incorporates Gaia-DR2 distance information \citep{2018A&A...616A...1G,2018A&A...616A...2L}.
A full discussion will be presented in a forthcoming paper and the results will be available as part of GALAH-DR3.
The ages and distances for the GALAH-MSTO
stars are computed with the BSTEP code \citep{2018MNRAS.473.2004S}.
BSTEP provides a Bayesian estimate of intrinsic stellar parameters
from observed parameters by making use of stellar isochrones.
For results presented in this paper, we use the PARSEC-COLIBRI stellar isochrones \citep{2017ApJ...835...77M}.
To select stars with reliable ages, we adopt the following selection function for MSTO stars,
\be
(3.2 < \log g < 4.1) \& (5000 <\; T_{\rm eff}/{\rm K}\;<7000). \label{equ:msto_select}
\ee

\section{Chemical evolution model with radial mixing} \label{sec:model}
One of the main purpose of a Galactic model is to predict the
joint distribution of all possible stellar observables for stars in the Milky Way.
Due to large spectroscopic surveys, the following set of observables
position ${\bf x}$, velocity ${\bf v}$, age $\tau$, iron abundance
$[{\rm Fe/H}]'$ and $\alpha$ abundance  $[\alpha/[{\rm Fe}]'$ are readily available
for a large number of stars. Hence the distribution function
we wish to seek is $p({\bf x},{\bf v},\tau,{\rm [Fe/H]},[\alpha/{\rm Fe}]|\theta)$, where
$\theta$ denotes the free parameters of the model (which we sometimes omit
to shorten the equation). Full list of parameters and their adopted
values is given in \autoref{tab:coeff}.
Our model is inspired and based on the extended distribution function
model proposed by \citet{2015MNRAS.449.3479S} but improves upon it by adding significant new features, e.g., the distribution of
$\alpha$ elemental abundances, a new prescription for velocity dispersion of stars. We also simplify certain aspects
of the \citet{2015MNRAS.449.3479S} model, e.g., the phase
space distribution is described by the Shu distribution function
\citep{1969ApJ...158..505S} instead of an action based quasi-isothermal distribution function.

For simplicity we assume the Galaxy to be axisymmetric, i.e.,
none of the Galactic properties depend on the azimuth coordinate.
Strictly speaking this is not true, an evidence for this is the
presence of non-axisymmetry structures like bars and spiral arms.
The azimuthal crossing time scale is quite small compared to the age of the Galaxy, hence, except for very young stars, axisymmetry should still a good
approximation for majority of the stars in the Milky Way.
Due to axisymmetry we work in cylindrical coordinates, and express the
phase space in following coordinates $R$, $\phi$, $z$, $v_R$, $v_{\phi}$
and $v_z$.

At any given time the stars are born out of ISM, which is made up of cold gas that is on circular orbit around the Galactic center.
Stars inherit the elemental composition and kinematic properties of the gas from which they are born.
Hence, the fundamental building block of our model is a group of stars
born at a lookback time $\tau$ with birth radius $R_b$.
The chemical evolution of the Galaxy will dictate how the
abundances vary as a function of $\tau$ and $R_b$. We denote
the chemical evolution of iron abundance by ${\rm [Fe/H]}(\tau,R_b)$ and that of $\alpha$ elemental abundance by $[\alpha{\rm Fe}](\tau,R_b)$. If ${\rm [Fe/H]}'$
and $[\alpha/{\rm Fe}]'$ are observed with uncertainties
$\sigma_{\rm [Fe/H]}$ and $\sigma_{[\alpha{\rm Fe}]}$,
the joint distribution of observables can be modelled as
\be
p({\bf x},{\bf v},\tau,{\rm [Fe/H]'},[\alpha/{\rm Fe}]')=
p({\bf x},{\bf v}|\tau,R_b) p(\tau,R_b) \times  \nonumber \\
\frac{\partial{{\rm [Fe/H]}(\tau,R_b)}}{\partial{R_b}} \times \nonumber\\
\mathcal{N}\left([{\rm Fe/H}]'|[{\rm Fe/H}](\tau,R_b),\sigma_{[\alpha{\rm Fe}]}\right) \times \nonumber\\
\mathcal{N}([\alpha/{\rm Fe}]'|[\alpha/{\rm Fe}](\tau,R_b),\sigma_{\rm [Fe/H]})
\ee
To also include stellar mass $m$ in the joint distribution,
the right hand side (RHS) of the above equation should be multiplied by the
initial-mass-function of stars $\xi(m)$.
This is important when we want to take the selection function of the survey into account and we postpone this discussion to \autoref{sec:selfunc}.

Having specified the full joint distribution, it is easy to
explore any other projections of this distribution, e.g., by integrating
over azimuth $\phi$ and velocities $v_{\phi}$, $v_R$ and $v_z$ we get
$p(R,z,R_g,\tau,{\rm [Fe/H]},[\alpha/{\rm Fe}])$.
In this paper we are interested in the distribution of $({\rm [Fe/H]},[\alpha/{\rm Fe}])$ at a given $R$ and $z$, and this is given by
\be
p({\rm [Fe/H]'},[\alpha/{\rm Fe}]'|R,z)=\frac{p(R,z,{\rm [Fe/H]'},[\alpha/{\rm Fe}]')}{p(R,z)}
\ee

\begin{table*}
\caption{Parameters of the chemical evolution model}
\begin{tabular}{lllll}
\hline
Description & symbol & value & \\
\hline
Solar Radius & $R_{\odot}$ & 8.0 & kpc &\\
Circular velocity at Solar Radius & $\Theta_{\odot}$ & 232.0 & km/s &\\
Gravitational Potential & $\Phi(R,z)$& MWPotential2014-galpy & & \\
\hline
Age of disc & $\tau_{\rm max}$ & 13.0 & Gyr & \\
Late time star formation rate decay constant & $\tau_{\rm fall}$ & 10.0 & Gyr & \\
Early time star formation rate rise constant & $\tau_{\rm rise}$ & 0.63 & Gyr & \\
\hline
Current metallicity gradient & $F_R$ & -0.08 & dex/kpc & \\
Metallicity at birth & $F_{\rm min}$ & -0.85 & dex & \\
Radius of current ISM solar metallicity & $r_F$ & 6.5 & kpc &\\
ISM Metallicity enrichment time scale & $\tau_F$ & 3.2 & Gyr & \\
\hline
$[\alpha/{\rm Fe}]$ transition time & $\tau_{\alpha}$ & 10.5 & Gyr &\\
Maximum $[\alpha/{\rm Fe}]$ & $\alpha_{\rm max}$ & 0.225 & dex & \\
Current $[\alpha/{\rm Fe}]$ of outermost disc& $\alpha_{\rm outer}$ & 0.1 & dex &\\
Transition metallicity & $F_{\alpha}$ & -0.5 & dex &\\
Transition metallicity scale & $\Delta F_{\alpha}$ & 0.5 & dex &\\
Time scale for transition of $[\alpha/{\rm Fe}]$ & $\Delta \tau_{\alpha}$ & 1.5 & Gyr & \\
\hline
Maximum radial scale length & $R_{d}^{\rm max}$ & 3.45 & kpc & \\
Minimum radial scale length & $R_{d}^{\rm min}$ & 2.31 & kpc & \\
Time of transition of radial scale length & $\tau_{R_d}$ & 9.0 & Gyr & \\
Time scale for transition of radial scale length & $\Delta \tau_{R_d}$ & 1.0 & Gyr & \\
\hline
Churning efficiency & $\sigma_{L0}$ & 1150 & kpc km/s & \\
\hline
Vertical Velocity dispersion normalization & $\sigma_{0,vz}$  & 25.0 & km/s & \\
Radial Velocity dispersion normalization & $\sigma_{0,vR}$  & 39.6 & km/s & \\
Vertical heating growth parameter & $\beta_{z}$   & 0.441 & \\
Radial heating growth parameter & $\beta_{R}$   & 0.251 & \\
Vertical heating angular momentum scale& $\lambda_{L,vz}$  & 1130  & kpc km/s & \\
Radial heating angular momentum scale& $\lambda_{L,R}$  & 2300 & kpc km/s & \\
Vertical heating angular momentum coefficient & $\alpha_{L,z}$  & 0.58 & \\
Radial heating angular momentum coefficient & $\alpha_{L,R}$  & 0.09 & \\
Vertical dispersion gradient with metallicity & $\gamma_{\rm [Fe/H],z}$ &  [-0.52,-0.8] & km/s/dex & \\
Radial dispersion gradient with metallicity & $\gamma_{\rm [Fe/H],R}$ &  [-0.19,-0.5] & km/s/dex & \\
\hline
\end{tabular}
\label{tab:coeff}
\end{table*}

\subsection{Phase space distribution and radial mixing} \label{sec:phasespace}
\begin{figure}[htb]
\centering \includegraphics[width=0.49\textwidth]{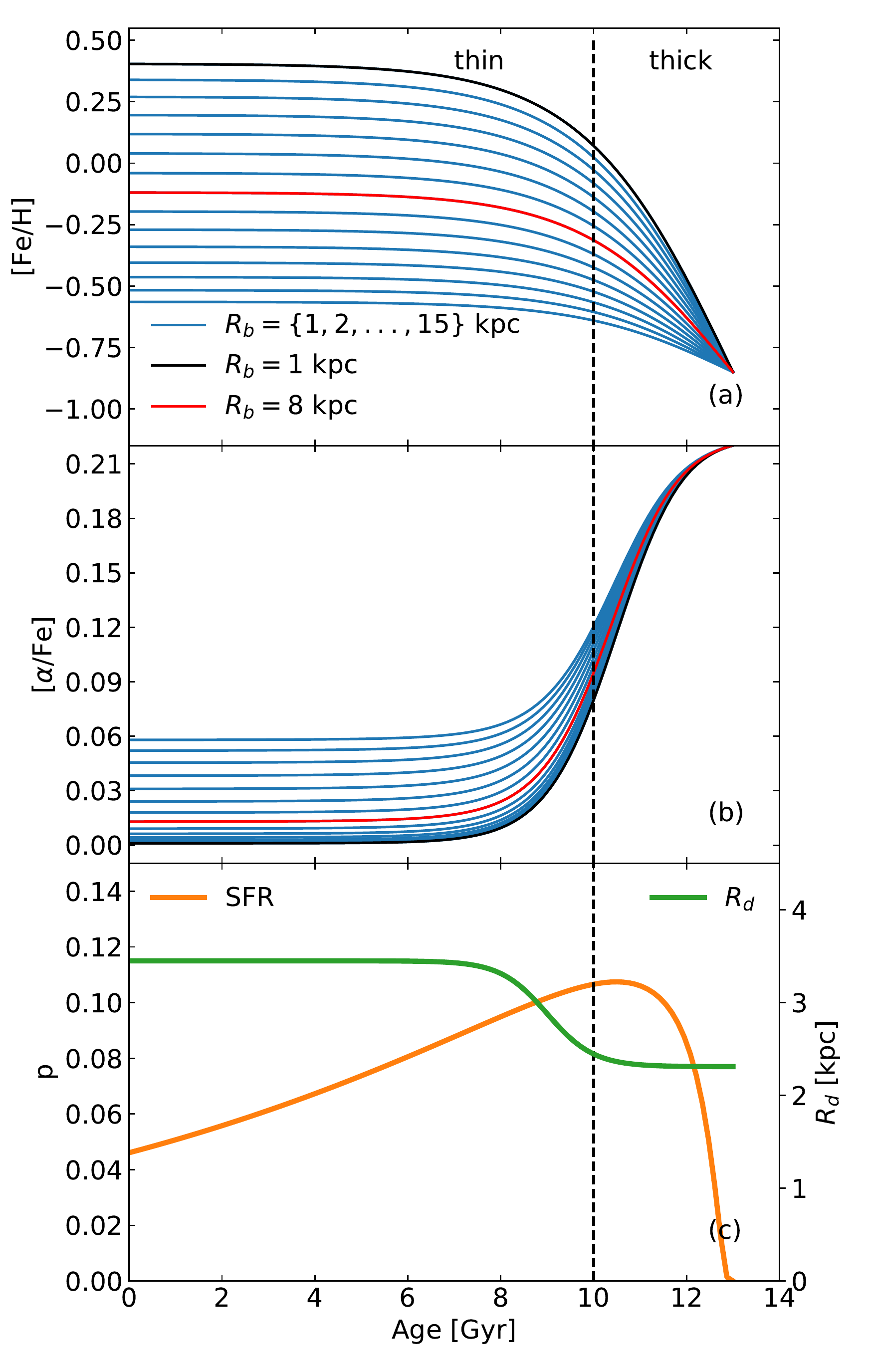}
\caption{Properties of our Galactic model. (a) [Fe/H] as a function of age for different birth radius. (b) [$\alpha$/Fe] as a function of age for different birth radius. (c) The star formation rate and radial scale length as function of age. The dotted lines simply show for reference the traditional definition of thick and thin disc based on age.
\label{fig:amr_ism}}
\end{figure}

The newly formed stars are also on circular
orbit with a star born at radius $R_b$ having an angular momentum
$v_{\rm c}(R_b)R_b$, where $v_{\rm c}(R_b)$ is the circular velocity.
The distribution of newly formed stars $p(\tau,R_b)$ is fully specified by
specifying the star formation history $p(\tau)$ and the distribution
of birth radius $p(R_b|\tau)$ for a given $\tau$.
\be
p(\tau,R_b) &=& p(\tau) p(R_b|\tau)
\ee
Following \citet{2015MNRAS.449.3479S} we express
the star formation history as
\be
p(\tau) &\propto& \exp\left(\frac{\tau}{\tau_{\rm fall}}-\frac{\tau_{\rm rise}}{\tau_{\rm max}-\tau}\right),
\ee
which is marked by a peak at
$\tau=\tau_{\rm max}-\sqrt{\tau_{\rm fall} \tau_{\rm rise}}$. For our choice of parameters the peak is at 10.5 Gyr, see \autoref{fig:amr_ism}c.
The star formation increases at earlier times with a rate controlled by
$\tau_{\rm rise}$ to a maximum value and then falls off exponentially
untill the present time with time scale $\tau_{\rm fall}$.

The radial distribution of stars at birth is given by
\be
p(R_b|\tau) &=& \frac{R_b}{R_d^2}\exp\left(-R_b/R_d\right),
\label{equ:rb_dist}
\ee
Unlike \citet{2015MNRAS.449.3479S}, who consider distinct thin
($\tau < 10$ Gyr) and thick ($\tau > 10$ Gyr) discs with different scale lengths, we allow for a smooth inside out formation of
the disc by specifying the scale length $R_d$ to evolve with time according to
\be
R_d &=& R_d^{\rm max}-\frac{R_d^{\rm max}-R_d^{\rm min}}{2}\left({\rm tanh}\left(\frac{\tau-\tau_{R_{d}}}{\Delta \tau_{R_d}}\right)+1\right),
\ee
the corresponding profile is shown in \autoref{fig:amr_ism}c.

Over time, due to various dynamical processes, like scattering from spiral arms, giant molecular clouds and a bar, stars move away from their place of birth and acquire random motion.
Following \citet{2009MNRAS.396..203S} we describe the dynamical processes using the churning and blurring mechanisms. Churning
refers to the scattering in angular momentum space, while blurring refers to increase of random motion that is characterized by radial velocity dispersion $\sigma_R$ and vertical velocity dispersion $\sigma_z$.
We assume $\sigma_R$ and $\sigma_z$ to be functions of
$\tau$, $R_b$, and $R_g$ (guiding radius, defined as the radius of a circular orbit with a given angular momentum ).

Specifically, due to churning stars born at a lookback time $\tau$ and at radius $R_b$ ,  will have a distribution of angular momentum $L$ or equivalently guiding radius $R_g(L)$ given by $p(R_g|\tau,R_b)$.
Following \citet{2015MNRAS.449.3479S} we model churning
as a Gaussian diffusion in the space of angular momentum $L$,
which leads to
\be
p(R_g|R_b,\tau) &=& \frac{1}{K} \mathcal{N}\left(L|R_b \Theta_{\odot}-\frac{\sigma^2_{L}}{2 \Theta_{\odot} R_d},\sigma_{L}^2\right)\frac{{\rm d}L}{{\rm d}R_g}.
\ee
Here $\sigma_{L}$ characterizes the dispersion of angular momentum
which increases with time according to
\be
\sigma_L(\tau)& =& \sigma_{L0} \left(\frac{\tau}{\tau_{\rm max}}\right)^{1/2}.
\ee
The distribution is only valid for positive values of $L$,
the factor
\be
K=\frac{1}{2}\left[1+{\rm erf}\left(\frac{R_b\Theta_{\odot}-\frac{\sigma^2_{L}}{2 \Theta_{\odot} R_d}}{\sqrt{2}\sigma_L}\right)\right]
\ee
is a normalization constant to ensure that
the integral over the positive $L$ axis is unity.

To model the present
day phase space distribution of stars born at a lookback time of $\tau$ and at radius $R_b$, $p({\bf x},{\bf v}|\tau,R_b)$, we use a distribution function of the following form \citep[see Equation 4.147 from][]{2008gady.book.....B}.
\be
f(E_R,L,E_z) \propto \frac{F(L)}{\sigma_R^2}\exp\left(-\frac{E_R}{2\sigma_R^2}\right)\exp\left(-\frac{E_z}{2\sigma_z^2}\right)\
\ee
Here, the potential $\Phi(R,z)$ is assumed to be linearly separable in $R$ and $z$ allowing the vertical and planar
motion to be studied separately. The planar distribution is modeled using
the Shu distribution function while the vertical distribution
is modelled as an isothermal population.
\be
E_z=\frac{1}{2}v_z^2+(\Phi(R,z)-\Phi(R,0))
\ee
is the energy associated with the vertical motion.
$\Phi(R,z)$ is the Galactic gravitational potential and we adopt the MWPotential2014 from galpy \citet{2015ApJS..216...29B}.
$E_R$ is the random energy over and above that of $E_c(L)$ (energy required for a star with a given $L$ to be in a circular orbit with radius $R_g(L)$) and is given by
\be
E_R&=&E-E_c(L)=\frac{V_R^2}{2}+\Phi_{\rm eff}(R,R_g)-\Phi_{\rm eff}(R_g,R_g) \nonumber \\
&=&\frac{V_R^2}{2}+\Delta \Phi (R,R_g).
\ee
$\Phi_{\rm eff}(R,R_g)$ is the effective potential for a planar orbit and is given by
\be
\Phi_{\rm eff}(R,R_g) &=& \Phi(R)+\frac{1}{2}v_{c}(R_g)^2(R_g/R)^2,
\ee
Given that we assume $\sigma_R$ to be a function of $\tau$, $R_b$, and $R_g$, the phase space distribution can now be written as
\be
p({\bf x},{\bf v}|\tau,R_b) &=& p(R_g|\tau,R_b)p(R|R_g,\sigma_R)  \times \nonumber \\ && \frac{1}{\sqrt{2\pi}\sigma_R} \exp\left(-\frac{v_{R}^{2}}{2\sigma_R^2}\right) \times \nonumber\\
&& p(z,v_z|R,\sigma_z) \frac{1}{2\pi}. \label{equ:churblur}
\ee

It follows from \citet{2013ApJ...773..183S} \citep[see also][]{2012MNRAS.419.1546S} that for a Shu distribution function,
\be
p(R|R_g,\sigma_R)&=&\frac{p(R,R_g|\sigma_R)}{\int P(R,R_g|\sigma_R) {\rm d}R} \nonumber \\
&=&\frac{1}{g_K(a,R_g)R_g} \times \\
&&\exp\left( -\frac{\Phi_{\rm eff}(R,R_g)-\Phi_{\rm eff}(R_g,R_g)}{\sigma_R^2}\right),
\ee
where $a=\sigma_R/\Theta(R_g)$, and
\be
g_K(a,R_g)=\int \exp\left( -\frac{\Phi_{\rm eff}(R,R_g)-\Phi_{\rm eff}(R_g,R_g)}{a^2v_{\rm c}^2(R_g)}\right) \frac{{\rm d}R}{R_g}
\ee

The vertical phase space distribution of stars at a given $R$ for an isothermal population characterized by vertical velocity dispersion $\sigma_z$ is given by
\be
p(z,v_z|R,\sigma_{z}) &=& \frac{1}{2 z_0} \exp\left(-\frac{\Phi(R,z)-\Phi(R,0)}{\sigma^2_{z}}\right) \times \\
&&\frac{1}{\sqrt{2\pi}\sigma_z} \exp\left(-\frac{v_{z}^{2}}{2\sigma_z^2}\right),
\ee
where $z_0$ is the vertical scale height \citep[see Equation 4.153 from][]{2008gady.book.....B} and is given by
\be
z_0(R,\sigma_{z})=\int_{0}^{\infty} \exp\left(-\frac{\Phi(R,z)-\Phi(R,0)}{\sigma^2_{z}}\right).
\ee

\begin{figure}[htb]
\centering \includegraphics[width=0.49\textwidth]{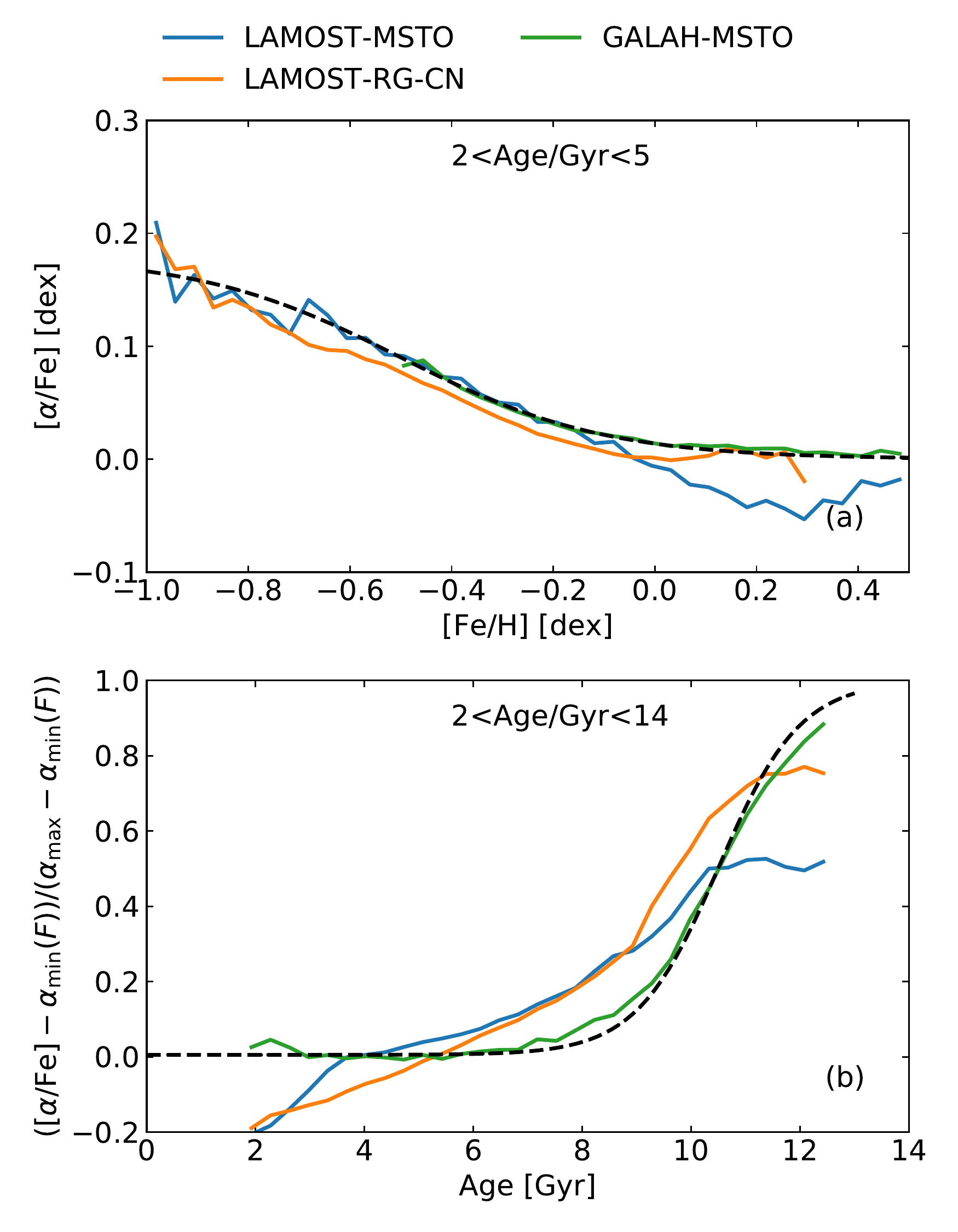}
\caption{Metallicity and age dependence of $\alpha$ elemental abundance. Solid lines show data from four different sources. The dashed lines are predictions of a model with, $\alpha_{\rm max}=0.225$, $\alpha_{\rm outer}=0.18$, $F_{\alpha}=-0.5$, $\Delta F_{\alpha}=0.4$, $\tau_{\alpha}=10.5$ Gyr and $\Delta \tau_{\alpha}=1.5$, which fits the GALAH data. $F$ stands for metallicity [Fe/H], and $\alpha_{\rm min}(F)$ is an analytical function of metallicity as shown by the dashed line in panel (a).
\label{fig:alpha_feh_profile}}
\end{figure}
\subsection{Chemical evolution}
The abundance of elements as a function of time and birth radius is dictated by the
chemical evolution of the Galaxy.
Chemical evolution from first principles involves tracking the birth and death of stars, the synthesis of elements in stars using nucleosynthetic yields, the return of synthesized elements to the ISM, the dilution of the ISM by the infall of fresh gas and so on. Rather than adopting an ab-into approach we adopt an empirical approach.
We specify simple but physically motivated functional forms for the evolution of abundances as a function
of $\tau$ and $R_b$ and fine tune some of the free parameters using observational data.

\subsubsection{Iron abundance}
For iron abundance [Fe/H], a reasonable assumption is to assume that it decreases monotonically with birth radius at all times. This is motivated by the fact that the star formation efficiency is highest in the center of the Galaxy and falls off with radius. A metallicity gradient of about $-0.09$ dex/kpc
has been observed in the Milky Way \citep{2014AJ....147..116H}. Most chemical evolution models also predict metallicity to fall off with birth radius \citep{2009MNRAS.396..203S}.
As for the dependence of metallicity on time, the models like that of \citet{2009MNRAS.396..203S} predict a sharp increase in metallicity at earlier times, but at later times
the rate of increase progressively slows down and the
metallicity approaches an asymptotic value $F_{\rm max}(R_b)$, which depends on $R_b$.
These above mentioned features are captured by the following adopted  functional form. Note, for clarity and brevity, $F$ is used to denote [Fe/H].
\be
&& F(R_b,\tau)=F_{\rm min}+(F_{\rm max}(R_b)-F_{\rm min}){\rm tanh}\left(\frac{\tau_{\rm max}-\tau}{\tau_F}\right) \label{equ:fe_tau}\\
&&F_{\rm max}(R_b)=F_{\rm min}{\rm tanh}\left(
\frac{F_R(R_b-R_F)}{F_{\rm min}}\right) \label{equ:fe_rb}
\ee
The $F_{\rm min}$ denotes the minimum ISM metallicity,
$\tau_F$ the metallicity enrichment time scale, $F_R$
the current metallicity gradient in the solar neighborhood,
$r_F$ the radius at where the ISM has solar metallicity.
\autoref{equ:fe_tau} is depicted graphically in \autoref{fig:amr_ism}a.
The model is similar to that of \citet{2015MNRAS.449.3479S},
except for the form of $F_{\rm max}(R_b)$--
our variation of radial gradient $dF_{\rm max}/dR_b$ with $R_b$ is weaker than that of \citet{2015MNRAS.449.3479S}.

\subsubsection{$\alpha$ elemental abundance}
For \alphafe{}, instead of expressing its formation
and evolution in terms of $\tau$ and $R_b$, we express it in terms
of $\tau$ and [Fe/H]. This is because
the birth radius cannot be observed directly, hence a relation
constructed out of $R_b$ is difficult to verify and calibrate.
However, using \autoref{equ:fe_tau} and \autoref{equ:fe_rb} we can
express $R_b$ in terms of [Fe/H] and $\tau$ as an analytical function
$R_b(F,\tau)$. This is possible because we assume that
[Fe/H] decreases monotonically with birth radius for any given age.

From previous works \citep{2017A&A...608L...1H,2017ApJS..232....2X},
\alphafe{} has been found to be approximately constant with age
till about 8 Gyr followed by a rapid rise thereafter.
We postulate a $\tanh$ function that transitions from
a low value $\alpha_{\rm min}$ to a high value $\alpha_{\rm max}$
at an age $t_{\alpha}$, with the sharpness of the transition
being controlled by $\Delta t_{\alpha}$.
\be
[\alpha/{\rm Fe}](F,\tau)&=&\alpha_{\rm min}(F)+ \nonumber \\ && \frac{\alpha_{\rm max}-\alpha_{\rm min}(F)}{2}\left[{\rm tanh}\left(\frac{\tau-\tau_{\alpha}}{\Delta \tau_{\alpha}}\right)+1\right]
\ee
The relationship is shown as dashed line in \autoref{fig:alpha_feh_profile}b.
The relationship is motivated by the physics of chemical
enrichment (Fe and $\alpha$ elements) in the Galaxy which
is mainly regulated by Supernovaes. The initial value
$\alpha_{\rm max}$ of \alphafe{}
is set by the yields of SNII, which occur almost
immediately (10 Myr) after the initiation of star formation at age $\tau_{\rm max}$. We expect $\alpha_{\rm max}$ to be independent
of metallicity $F$.
SNIa mostly produce Fe and almost no $\alpha$ elements,
which leads to a drop in \alphafe{}. SNIa require a
binary companion and can only occur after significant
time delay. The SNIa rates typically peak about 1 Gyr
after star formation. This typically sets the time
scale $\Delta \tau_{\alpha}$ of transition from high
to low \alphafe{}. We expect $\tau_{\alpha}$ to be given by
$\tau_{\rm max}-k\Delta \tau_{\alpha}$, with
$k$ being somewhere between 1 and 2, the exact value
needs to be determined by fitting to observational data.
As the evolution proceeds at some stage the ISM will
reach an equilibrium state due to infall of fresh metal poor gas
and this will set the floor $\alpha_{\rm min}$.
Since, the star formation rate and the infall rate
are not same at all birth radius, $\alpha_{\rm min}$ will
depend on birth radius. Given $F$ is a function of $R_b$
and $\tau$, we expect $\alpha_{\rm min}$ to be a function of $F$.

Given that \alphafe{} is approximately constant for young stars,
we can easily deduce the dependence of \alphafe{} on [Fe/H] for them,
and this is shown in \autoref{fig:alpha_feh_profile}a using
different spectroscopic data sets.
For young stars \alphafe{} is strongly anti-correlated with metallicity for $-0.8<{\rm [Fe/H]}<0$, but outside this range the slope approaches zero. We use the $\tanh$ function
\be
\alpha_{\rm min}(F)=\frac{\alpha_{\rm outer}}{2}\left[{\rm tanh}\left(-\frac{(F-F_{\alpha})}{\Delta F_{\alpha}} \right)+1\right]
\ee
to describe this relationship. $\alpha_{\rm outer}$ indicates the
\alphafe{} for young stars in the outer disc which have the least value of [Fe/H].
\autoref{fig:alpha_feh_profile}a shows that different observational data sets are all consistent with the adopted relationship. The data sets used are,
main sequence turnoff (MSTO) stars from the LAMOST survey, the
red-giant-branch (RGB) stars from the LAMOST survey, and
the MSTO stars from  the GALAH survey.

For older stars, the variation of age-\alphafe{} relation with metallicity is difficult to study, this is because old stars are mostly metal poor, which means it is difficult to get a sample with a wide range of metallicities. At earlier times, we expect the \alphafe{} to
be same throughout the disc as they are formed out of the
same primordial gas. Hence, we postulate $\alpha_{\rm max}$
to be independent of $F$. Additionally, we also postulate
$\tau_{\alpha}$ and $\Delta \tau_{\alpha}$ to be
independent of $F$.
\autoref{fig:alpha_feh_profile}b shows the observed dependence of
$([\alpha/{\rm Fe}]-\alpha_{\rm min}(F))/(\alpha_{\rm max}-\alpha_{\rm min}(F))$ on \feh for stars belonging to different data sets. GALAH-MSTO data set is consistent
with the adopted functional form and shows the
sharpest transition compared to other data sets, most likely due to better age precision.
For GALAH-MSTO, the relationship is very flat for stars younger than 8 Gyr, but for other data sets a small increase with age
can be seen.
Given that $R_b$ can be estimated from [Fe/H]
and $\tau$, we can now express $[\alpha/{\rm Fe}]$ in terms
of $\tau$ and $R_b$ and this is shown in \autoref{fig:amr_ism}b.
A detailed study of \alphafe{} as a function of age and
metallicity, based on the GALAH survey, will be presented in a forthcoming paper, we here adapt some of its relevant findings.
Due to systematic differences between spectroscopic surveys we need
to adjust the relations depending upon the survey we want to use it for.
The actual values that we use for building our model for APOGEE
data are given in \autoref{tab:coeff}, which differ slightly in values
for $\alpha_{\rm outer}$, $\Delta F_{\alpha}$ as compared to those given in \autoref{fig:alpha_feh_profile}.

\begin{figure*}[tb]
\centering \includegraphics[width=0.99\textwidth]{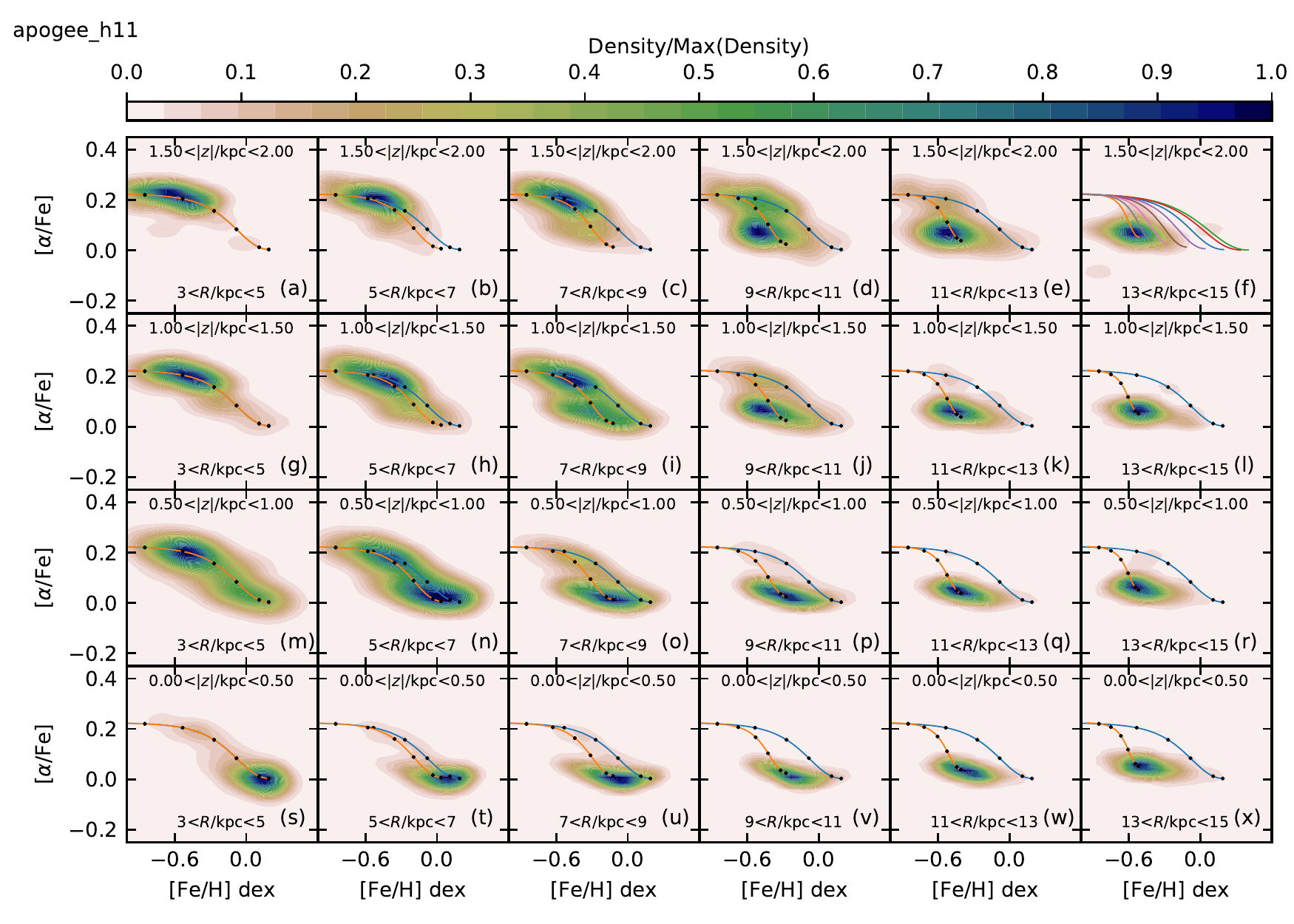}
\caption{Distribution of APOGEE stars in the $({\rm [Fe/H]},[\alpha/{\rm Fe}])$ plane at different locations in the Galaxy. The stars follow the selection function given in \autoref{equ:rgselect1}.
In each panel the density is normalized such that maximum density is unity.
The locations are specified in terms of
cylindrical coordinates $R$ and $z$ and the quoted values are in units of kpc. Each panel corresponds to a bin in $(R,|z|)$ space, with
$R$ increasing from left to right and $|z|$ increasing from bottom to top. In each panel
the solid lines show the evolution of abundances at a given birth radius taken from our model. The blue line is for the birth radius of 4 kpc, while the orange line is for the birth radius corresponding to the central value of $R$ in each bin. The black dots mark the evolution at age of 4, 8, 10, 11, 12, and 13 Gyr, with metallicity decreasing with age. In panel (f),
the model profiles corresponding to birth radii of 1, 2, 4, 6, 8, 10, 12 and 14 are shown, with [$\alpha$/Fe] increasing with birth radius.
\label{fig:apogee_data}}
\end{figure*}

\begin{figure*}[tb]
\centering
\includegraphics[width=0.99\textwidth]{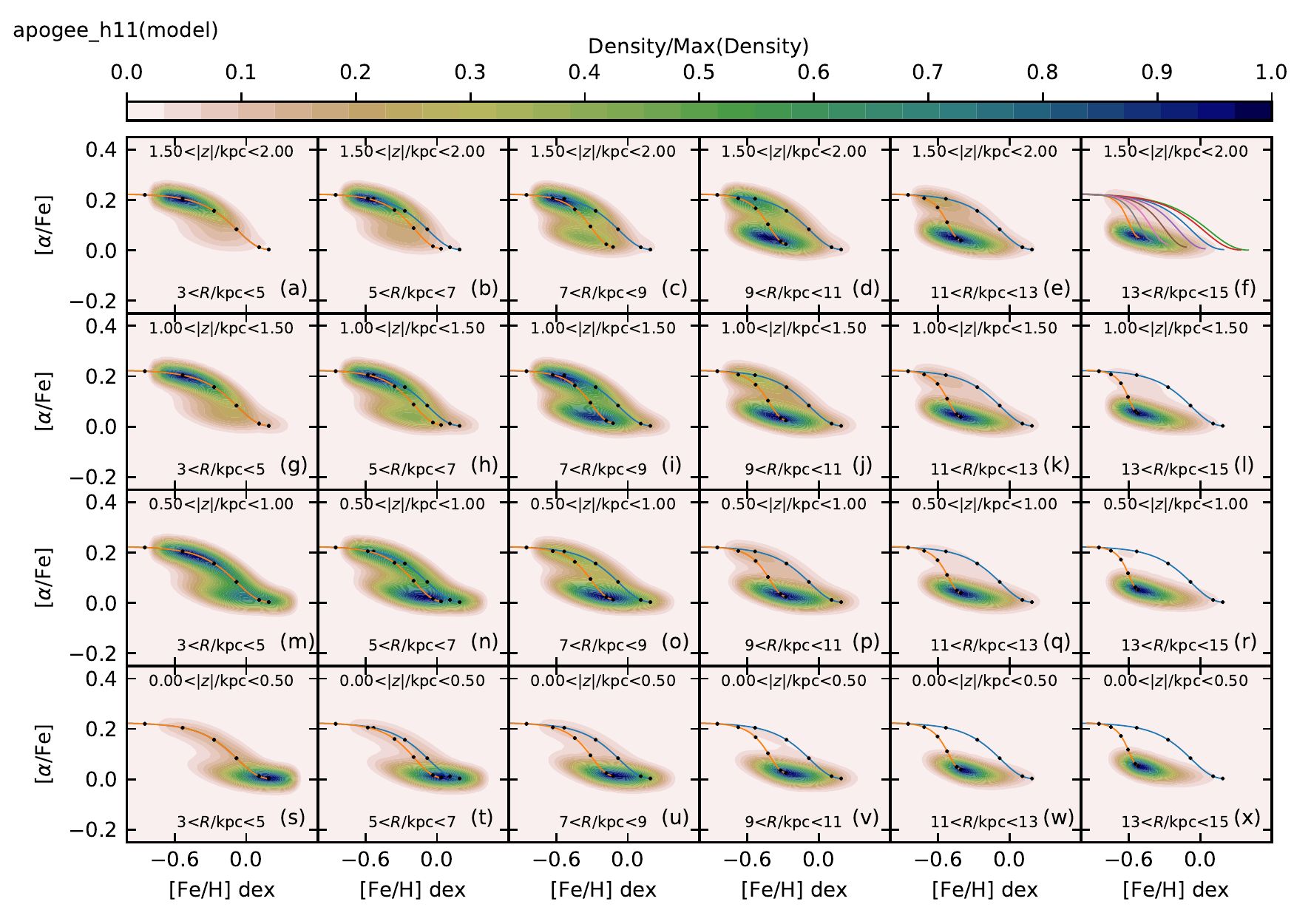}
\caption{Model predictions for distribution of stars in the $({\rm [Fe/H]},[\alpha/{\rm Fe}])$ plane at different locations in the Galaxy, satisfying the selection function of APOGEE stars (\autoref{equ:rgselect1}). Solid lines are evolutionary tracks for a given birth radius, for further description see \autoref{fig:apogee_data}
\label{fig:apogee_model}}
\end{figure*}

\begin{figure*}[tb]
\centering \includegraphics[width=0.99\textwidth]{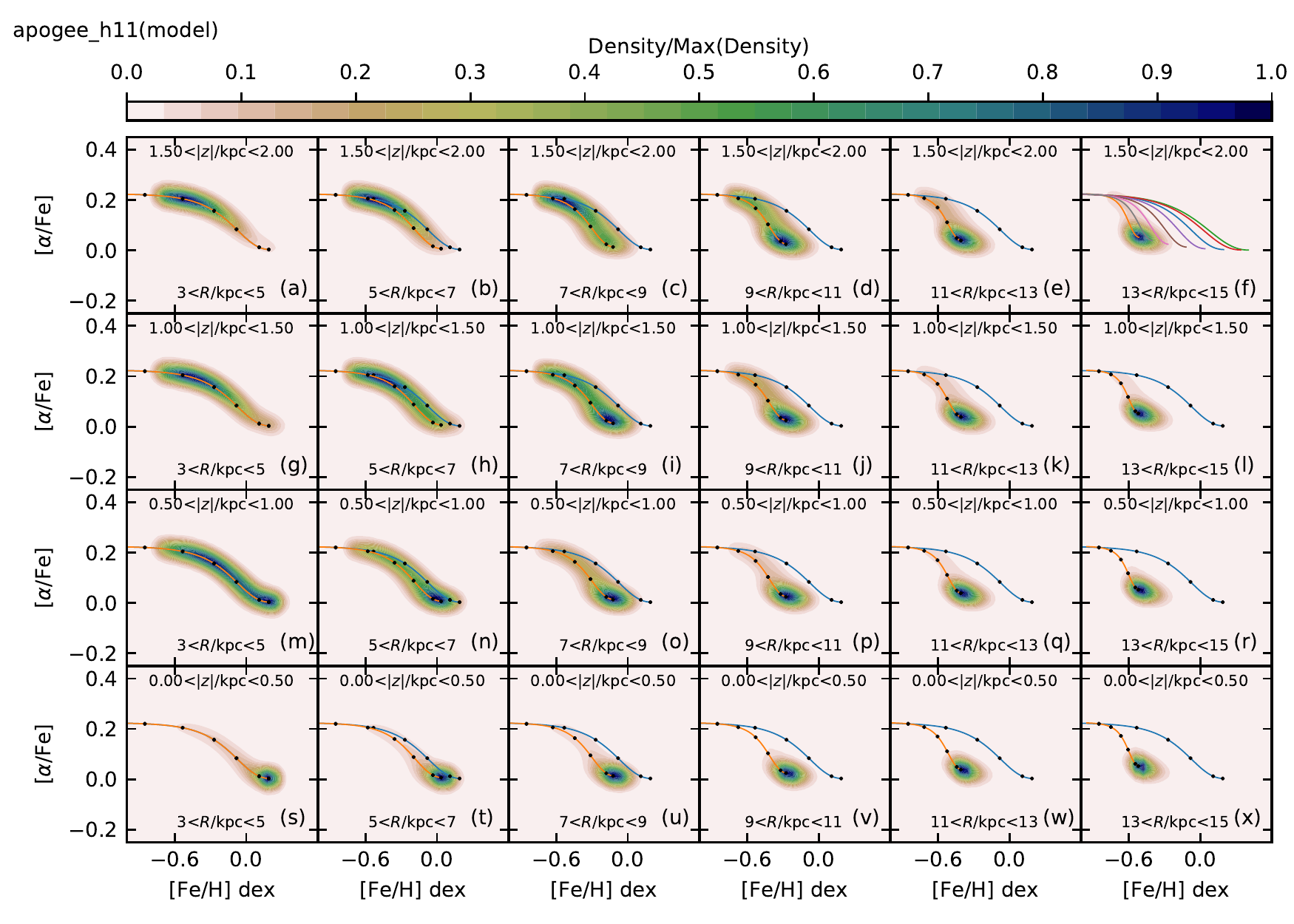}
\caption{Distribution of mock APOGEE stars in the $({\rm [Fe/H]},[\alpha/{\rm Fe}])$ plane at different locations in the Galaxy as predicted by a model with negligible churning ($\sigma_{L0}=150$ kpc km/s). The solid lines mark model evolutionary tracks for a given birth radius as described in \autoref{fig:apogee_data}. Abundances follow the profile corresponding to that of the local radial coordinate.
\label{fig:apogee_model_no_churn}}
\end{figure*}

\begin{figure*}[tb]
\centering \includegraphics[width=0.99\textwidth]{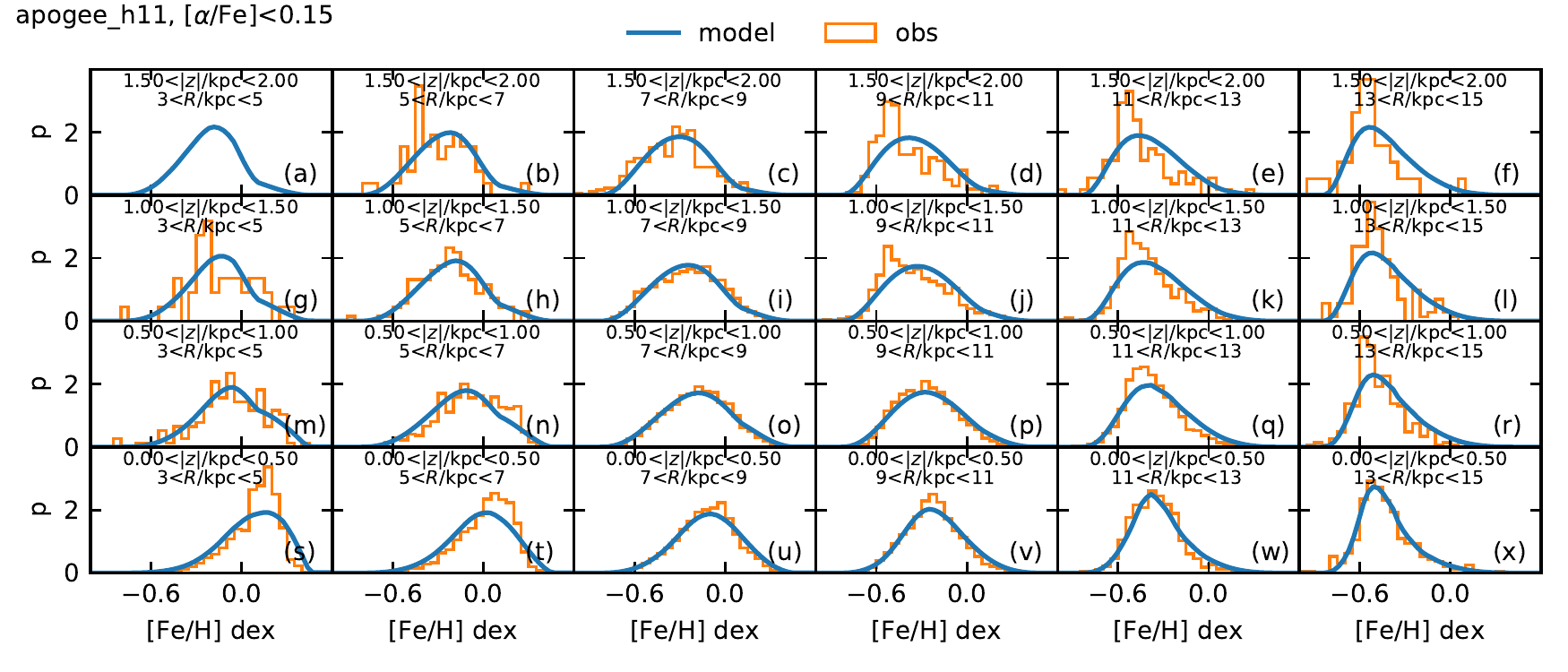}
\caption{Distribution of [Fe/H] for low-\alphafe{} stars from APOGEE along with predictions from our model.
\label{fig:p_fe_Rz_0}}
\end{figure*}
\begin{figure*}[tb]
\centering \includegraphics[width=0.99\textwidth]{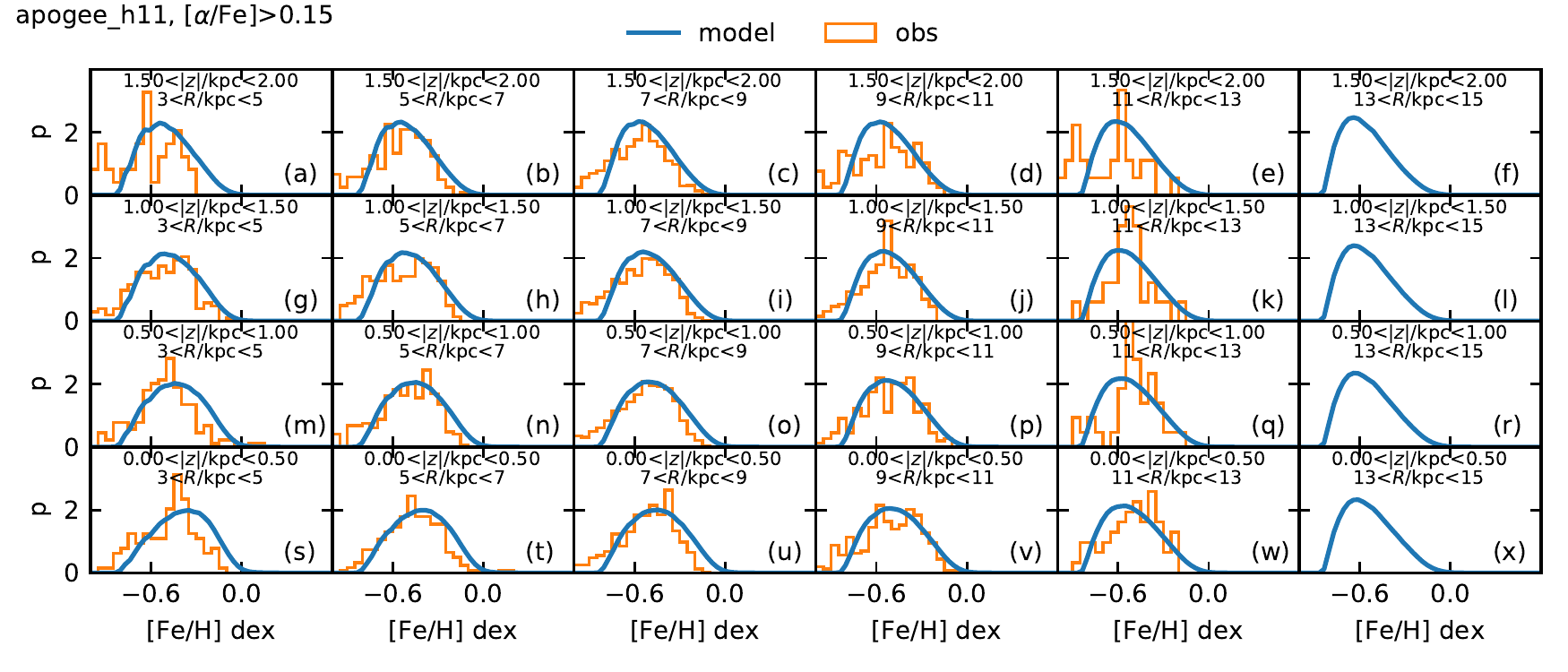}
\caption{Same as \autoref{fig:p_fe_Rz_0} but for high-\alphafe{} stars.
\label{fig:p_fe_Rz_1}}
\end{figure*}

\begin{figure*}[tb]
\centering \includegraphics[width=0.98\textwidth]{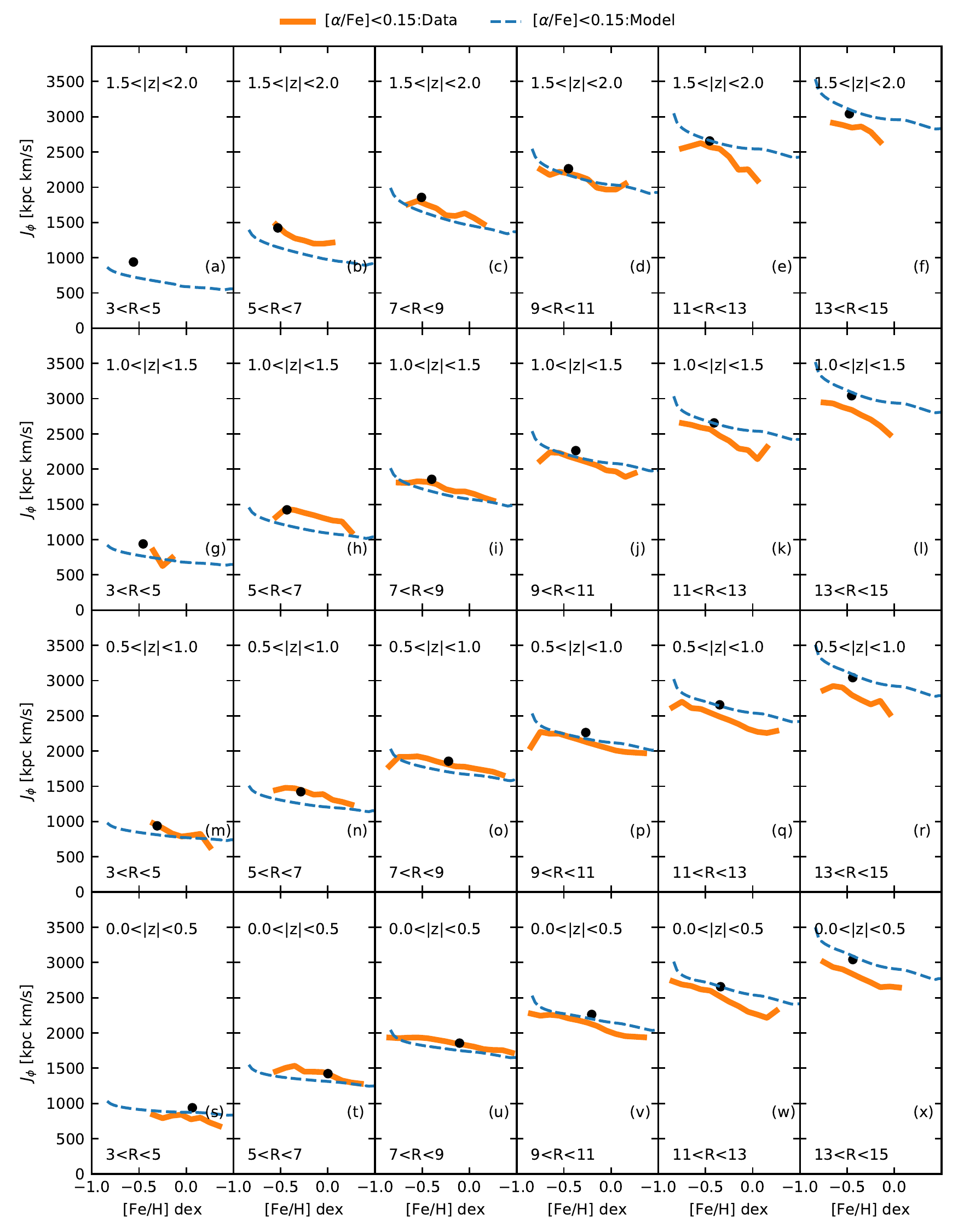}
\caption{Specific angular momentum as a function of [Fe/H], for low-\alphafe{} stars. Solid lines show observational data from LAMOST and APOGEE surveys. The dashed lines are model predictions. Different panels correspond to different locations in the Galaxy, specified by Galactocentric coordinates $R$ and $z$. The black dot marks the mean metallicity and angular momentum of a circular orbit corresponding to the mean radius of the stars in each $(R,z)$ bin.
\label{fig:Lz_lamapo_model}}
\end{figure*}

\begin{figure}[tb]
\centering \includegraphics[width=0.49\textwidth]{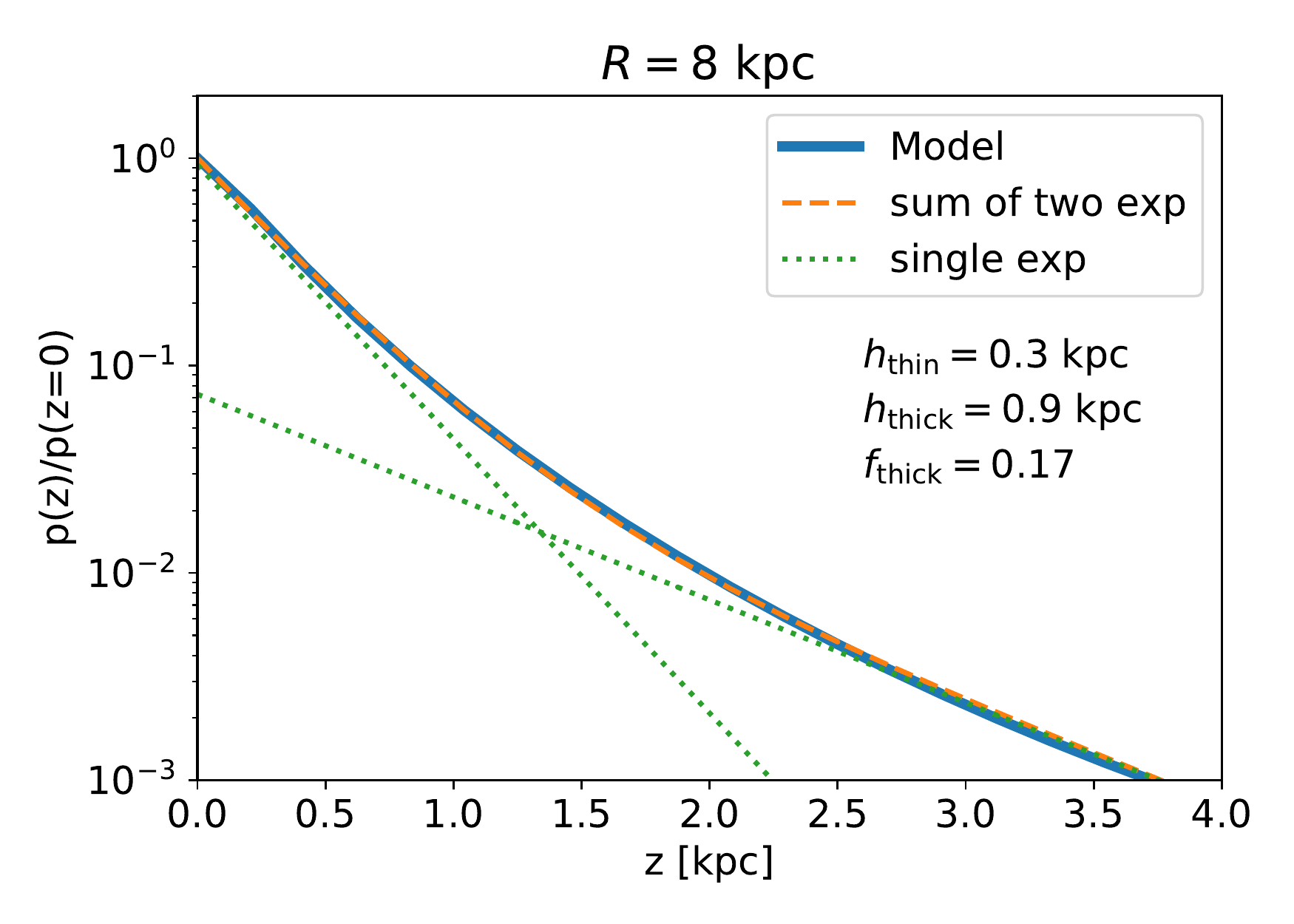}
\caption{Vertical distribution of star forming mass at $R=8.0$ kpc according to our model. The distribution is well fit by function that is a sum of two exponential functions, with scale lengths $h_{\rm thin}$ and $h_{\rm thick}$, and  fractional contribution of the thicker component $f_{\rm thick}$ being 0.17.
\label{fig:pz_R8}}
\end{figure}

\subsection{Velocity dispersion relations}
In \autoref{sec:phasespace} velocity dipsersions $\sigma_R$ and $\sigma_z$
were assumed to be a function of $\tau$, $R_b$ and $R_g$ (or equivalently  $L$). Given that $R_b$ can be expressed in terms of $\tau$ and [Fe/H], we
seek functions of the following form $\sigma_v(\tau,L,{\rm [Fe/H]})$.
Functions of the above form were explored by  \citet{2020arXiv200406556S}
using data from the LAMOST, GALAH and APOGEE spectroscopic surveys.
They showed that different stellar samples, even though they target different tracer populations and employ a variety of age estimation techniques, follow the same set of fundamental relations.
In addition to the well known power law dependence on age,
velocity dispersion is a parabolic shaped function of $L$ with a minima at around
solar angular momentum and it is anti-correlated with metallicity.

In \citet{2020arXiv200406556S}, the dispersion $\sigma_v$ of velocity $v$ (for either $v_R$ or $v_z$), is assumed to depend on the stellar
age $\tau$, angular momentum $L$, metallicity \feh{}, and vertical height from the disc midplane $z$, via the following multiplicatively separable functional form
\begin{equation}
\sigma_v(\tau,L_z,{\rm [Fe/H]},z,\theta_v)=\sigma_{0,v} f_{\tau}f_{L_z}f_{{\rm [Fe/H]}} f_{z},
\label{equ:vdisp_model}
\end{equation}
with
\begin{equation}
f_{\tau}=\left(\frac{\tau/{\rm Gyr}+0.1}{10+0.1}\right)^{\beta_v},
\label{equ:f_tau}
\end{equation}
\begin{equation}
f_{L}=\frac{\alpha_{L,v} (L/L_{\odot})^2+\exp[-(L-L_{\odot})/\lambda_{L,v}]}{1+\alpha_{L,v}},
\label{equ:f_lz}
\end{equation}
\begin{equation}
f_{\rm [Fe/H]}=1+\gamma_{{\rm [Fe/H]},v} {\rm [Fe/H]},
\label{equ:f_feh}
\end{equation}
\begin{equation}
f_{z}=1+\gamma_{z,v} |z|.
\label{equ:f_z}
\end{equation}
$\theta_v=\{\sigma_{0,v},\beta_{v},\lambda_{L,v},\alpha_{L,v},\gamma_{{\rm [Fe/H]},v},\gamma_{z,v}\}$ is a set of free parameters and we adopt the
values from \citep{2020arXiv200406556S} (also listed in \autoref{tab:coeff}).
The $\sigma_{0,v}$ is a constant that
denotes the velocity dispersion for stars lying in the midplane with solar metallicity, solar angular momentum ($L_{\odot}=\Omega_{\odot}R_{\odot}^2$) and an age of 10 Gyr.
Since in our models velocity dispersions have no dependence on $z$,
we set $\gamma_{z,v}$ to zero, and to compensate for it we increase $\sigma_{0z}$ by a few km/s.
The origin of the $z$ dependence is not fully understood, and more work is required in future before we can successfully incorporate them in theoretical models. Based
on results of \citet{2020arXiv200406556S} where $\gamma_{{\rm [Fe/H]},v}$ was found to decrease with age, we allow it to vary linearly with
age and the adopted maximum and minimum values are given in \autoref{tab:coeff}.

\subsection{Selection Function} \label{sec:selfunc}
To compare the predictions of a model with observations, we need to take the selection function of the observational data into account.
Let $S$ denote the event that a star is in a survey
based on criteria defined over some set of observables ${\bf y}$, e.g. $\log g$, $T_{\rm eff}$, apparent magnitude $H$.
The selection function of a survey
$p(S|{\bf y})$ is then the probability of
the event $S$ given ${\bf y}$ \citep{2012MNRAS.427.2119S}.
This is typically an indicator function that
is 1 if the star satisfies the selection criteria and 0 if it does not. Given the intrinsic variables
age $\tau$, metallicity \feh{}, distance $s$ and mass $m$, one can predict any observable ${\bf y}$ using
theoretical stellar isochrones. Hence the
selection function can also be computed over the
intrinsic variables.
For a given initial-mass-function $\xi(m)$ (IMF), normalized such that
$\int \xi(m)dm=1$, we have
\be
p(S|\tau,{\rm [Fe/H]},s)=\int p(S|\tau,{\rm [Fe/H]},s,m) \xi(m) dm.
\ee
We use PARSEC-COLIBRI stellar isochrones \citep{2017ApJ...835...77M} to compute this.

In this paper, we are mainly interested in
the distribution of \feh{} and \alphafe{} for stars
in a bin $k_{Rz}$ in $(R,|z|)$ space and chosen with some given selection function $S$. This required distribution is given by
\be
p({\rm [Fe/H]}, [\alpha/{\rm Fe}] | S, k_{Rz}) &=&
\int  p({\rm [Fe/H]}, [\alpha/{\rm Fe}], \tau | S, k_{Rz}) {\rm d}\tau. \nonumber \\
&&
\ee
Assuming that the bin $k_{Rz}$ is small enough such that $p({\rm [Fe/H]}, [\alpha/{\rm Fe}], \tau | R, z)$ is constant over the bin, we have
\be
p({\rm [Fe/H]}, [\alpha/{\rm Fe}], \tau | S, k_{Rz}) &=&
p({\rm [Fe/H]}, [\alpha/{\rm Fe}], \tau | R, z) \times \nonumber \\
&& p(S | \tau, {\rm [Fe/H]}, k_{Rz}),
\ee
with
\be
p(S|\tau,{\rm [Fe/H]},k_{Rz})= \int p(S|\tau,{\rm [Fe/H]},s) p(s|k_{Rz}){\rm d}s,
\ee
and $p(s|k_{Rz})$ being the distribution of distances of observed stars in bin $k_{Rz}$.

\section{Results}
We explore the joint distribution of \feh{} and \alphafe{}
at different $R$ and $|z|$ locations in the Galaxy. First, we present
observational results. Next, we compare the observational results
with the predictions of our theoretical model from \autoref{sec:model}.
\autoref{fig:apogee_data} shows the observational results from APOGEE-DR14 survey. This was first presented by \citet{2015ApJ...808..132H},
our figure here is a reproduction of their figure but with a few changes.
Our $(R, |z|)$ grid is slightly different, we have an extra bin in $|z|$. Also our target
selection criteria is more conservative, so that the selection
function can be easily reproduced when we do forward modelling of the
observed data. In \autoref{fig:apogee_data}, to aid comparison with theoretical predictions, the chemical evolutionary tracks corresponding to different birth radius
are indicated by solid lines.
Black dots mark the progression of time,
with [Fe/H] increasing (decreasing) with time (age).
The blue line is for $R_b=4$ kpc, while the orange line
is for $R_b$ corresponding to the central value of $R$ in each panel.

In \autoref{fig:apogee_data}, a double sequence (high-\alphafe{} and low-\alphafe{}) is visible in most panels, e.g., panels (d), (i), (m), (n) and (o). The two sequences are
well separated at the low [Fe/H] end, but with increase of [Fe/H] they progressively approach each other and eventually merge at \feh{} of about 0.
The relative number of stars belonging to each sequence depends
sensitively upon $R$ and $|z|$. The fraction of stars
belonging to the high-\alphafe{} sequence increases with increase of height
$|z|$ and decrease of radius $R$, in other words, the fraction is strongest away from the disc plane  and towards the inner disc (top-left panel).
The opposite is true for the low-\alphafe{} sequence, which is strongest
close to the disc plane and towards the outer disc (bottom-right panel). The high-\alphafe{} sequence appears to follow the track with $R_b=4$ kpc in all panels,  and the distribution of stars along this track is also very similar in all panels-- a property that we refer to as uniformity of the high-\alphafe{} sequence. In contrast, the distribution of stars along the low-\alphafe{} sequence is highly variable. With increase of either $R$ (going from the inner disc to the outer disc) or $|z|$ (going from midplane upwards), the density peak shits to the left, i.e., towards lower values of [Fe/H].

\autoref{fig:apogee_model} shows the distribution of stars in the \fehalpha{} plane predicted by
our model for the same $(R,|z|)$ grid as in \autoref{fig:apogee_data}
and using the same target selection criteria (\autoref{equ:rgselect1}). The predicted distributions
are strikingly similar to the observed distributions and are even found to
reproduce some of the finer details of the observed distributions.
Some examples of similarities are as follows.
The double sequence is prominent in panels (d), (i), (m), (n) and (o).
The relative fraction of stars in the two sequences varies with $R$
and $|z|$ in the same way as in \autoref{fig:apogee_data}.
The high-\alphafe{}
sequence is strongest at high $|z|$ and small $R$, and gradually
diminishes in strength with increase of $R$ and decrease of $|z|$.
The high-\alphafe{} sequence seems to follow the $R_b=4$ kpc evolutionary track in all panels.
For the low-\alphafe{} sequence, the [Fe/H] and \alphafe{} coordinates of the density peak change  with $R$ and $|z|$ in exactly the same way as in \autoref{fig:apogee_data}. To summarize, \autoref{fig:apogee_model}
demonstrates that our chemodynamical model can successfully reproduce the
observed distribution of stars in the \fehalpha{} plane across
different locations in the Galaxy.

\autoref{fig:apogee_model_no_churn} shows the \fehalpha{} distribution predicted by our model where churning is set to be negligible,
and it looks very different from \autoref{fig:apogee_model}. Unlike the double sequence seen in \autoref{fig:apogee_model},  only one sequence can be seen in \autoref{fig:apogee_model_no_churn}. In each panel, the sequence mainly follows the orange line, which is the evolutionary track with $R_b$ equal to
mean radius $R$ of each panel. Close to the plane, the sequence
is more like a blob which is centered around the black dot corresponding to
age of 4 Gyr. However, with increase of $|z|$ the sequence becomes elongated
and moves upward towards older stars. This is because the scale height increases with age (due to increase of $\sigma_z$ with age), which makes it more likely to have old stars at higher $|z|$. \autoref{fig:apogee_model_no_churn} makes it clear that radial migration, or more precisely the process of churning, is essential to get the double $\alpha$-sequence.
Note, blurring was kept unchanged and its effect is included in \autoref{fig:apogee_model_no_churn}. So blurring by itself is not enough to bring
stars to a given $R$ from a birth radius that is too far from $R$.

An important observation made by \citet{2015ApJ...808..132H} was that the shape of
\feh{} distribution (MDF) changes systematically with $R$, for stars close to the plane
the skewness changes from being negative in the inner disc ($R<7$ kpc) to being
positive in the outer disc ($R>11$ kpc). To investigate this, we split the APOGGE data set
used in \autoref{fig:apogee_data} into low and high $\alpha$ sample, and show with orange lines,
the observed MDFs for the low-\alphafe{}  and the high-\alphafe{} stars in \autoref{fig:p_fe_Rz_0} and \autoref{fig:p_fe_Rz_1} respectively. The MDFs are shown at different $R$ and $|z|$ locations.
The predictions of the model are shown alongside as blue lines.
For low-\alphafe{} stars, overall the model predictions are in very good agreement with the observations. Some panels show slight
differences, with observations showing sharper peak, e.g., panels (d), (e), (k) and (s).
For high-\alphafe{} stars, the model predictions are also in good
agreement with observations, however, in some panels the low metallicity tail is more extended in the observations.

An important prediction of the radial migration model is that along the low-\alphafe{} sequence the mean  specific-angular-momentum of stars should decrease systematically from the low-[Fe/H] end to the high-[Fe/H] end.
This is because, at any given $R$,  stars that have migrated from the inner disc carry less angular momentum than stars that have migrated from the
outer disc.
The model predictions are shown in \autoref{fig:Lz_lamapo_model}, where the angular momentum
is plotted as a function of \feh{} (dashed light-blue line) for stars belonging to the low-\alphafe{} sequence  for different $R$ and $|z|$ locations.
For panels with $R>5$ kpc, strong anti-correlation can be seen.
The relationship between $L$ and [Fe/H] is not universal, it varies
with $R$ and $|z|$. As expected, the mean angular momentum increases with $R$
in proportion to $v_{\rm circ}(R)R$, which is indicated by black dot.
The steepness of the profiles increases with $R$ and $|z|$.
Observations are also shown alongside (solid orange lines) which match reasonably well with the predictions.

The vertical distribution of star in the Milky Way is well fit by a sum of two exponential functions \citep{1983MNRAS.202.1025G}, leading to the suggestion that the Milky Way is made up of two distinct components the thin disc (with smaller scale height) and the thick disc (with larger scale height).
Model predictions for vertical distribution of stellar mass is shown in \autoref{fig:pz_R8}.
It can be seen that it is well fit by a sum of two exponential functions, although the model
does not have a distinct thick disc component. In our tests, a model with constant star formation rate and
scale length was also well fit by a sum of two exponential functions, but with slightly different fit parameters. To conclude, a continuous stars formation history can give rise to  vertical density distribution that is well fit by a sum of two exponential functions. A similar argument against the existence of a distinct thick disc was also presented by \citet{1987ApJ...314L..39N}.

\begin{figure*}[tb]
\centering \includegraphics[width=0.99\textwidth]{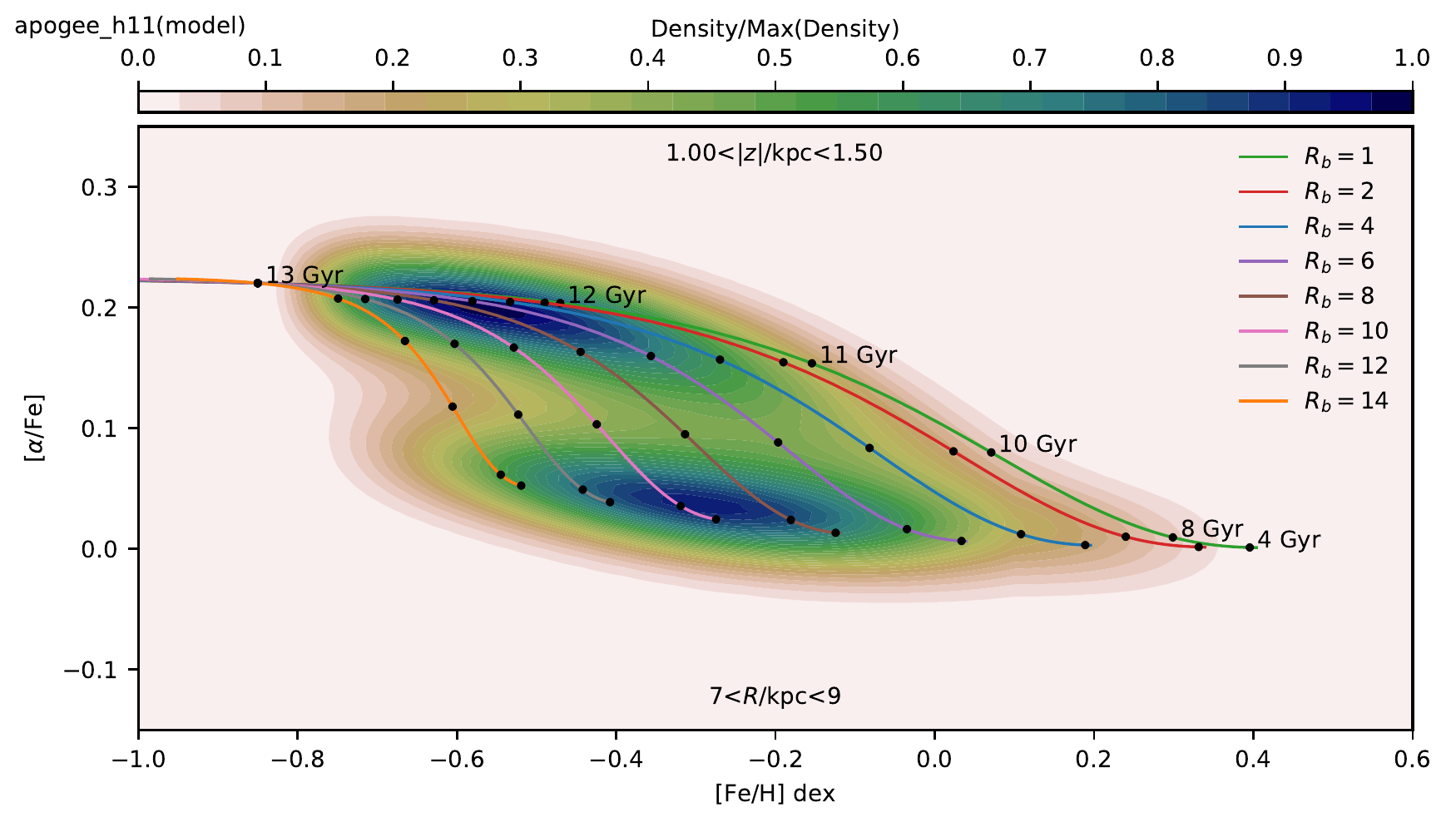}
\caption{Model prediction for the distribution of stars in the $({\rm [Fe/H]},[\alpha/{\rm Fe}])$ plane for $(7<R/{\rm kpc}<9)\ \& \ (1<|z|/{\rm kpc}<1.5)$.
The solid lines show the evolution of abundances for different birth radius $R_b$. The black dots mark the evolution at age of 4, 8, 10, 11, 12 and 13 Gyr.
\label{fig:alpha_fe_Rz_one}}
\end{figure*}

\begin{figure}[tb]
\centering \includegraphics[width=0.49\textwidth]{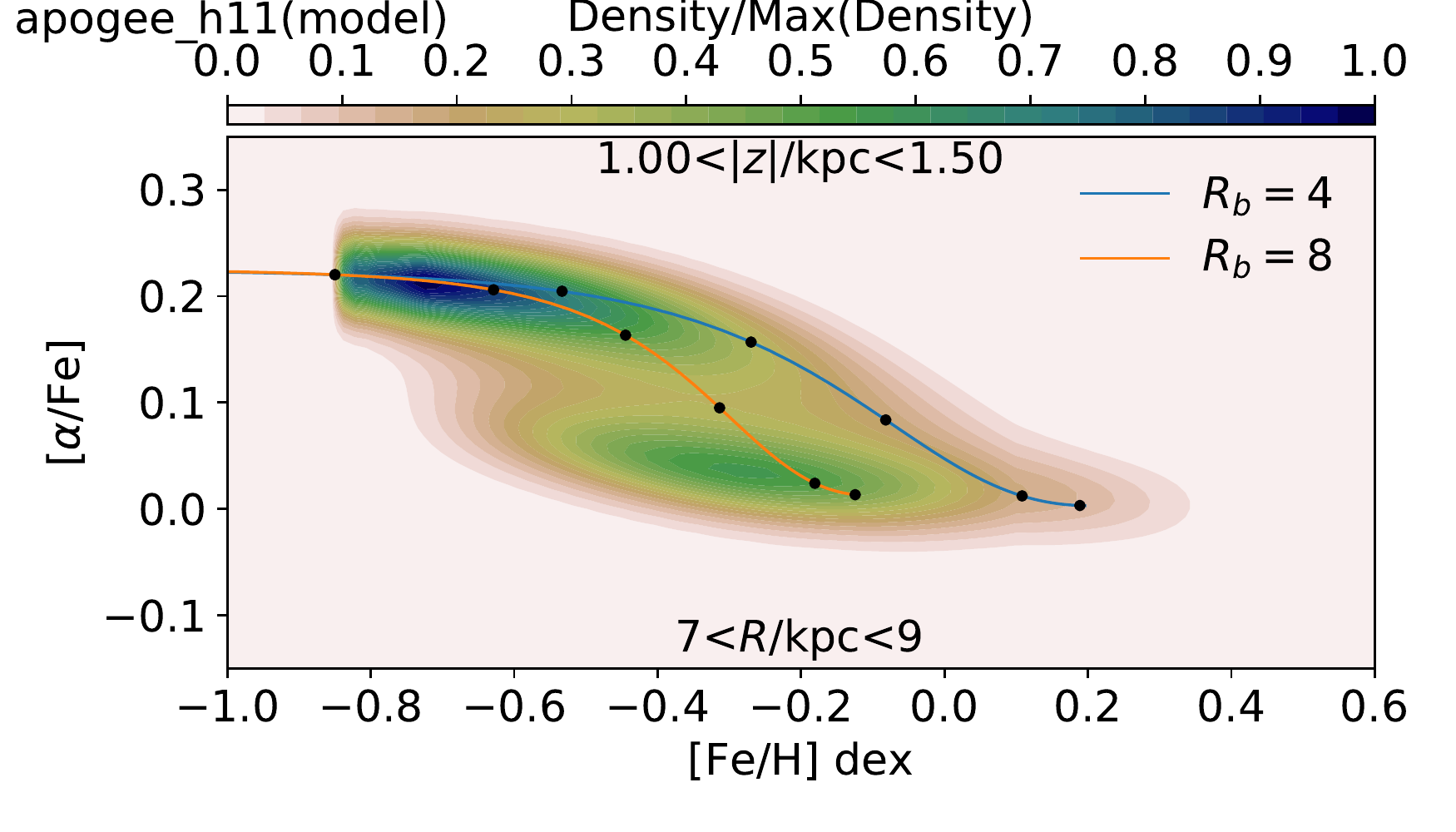}
\caption{Model prediction for the distribution of stars in the $({\rm [Fe/H]},[\alpha/{\rm Fe}])$ plane for $(7<R/{\rm kpc}<9)\ \& \ (1<|z|/{\rm kpc}<1.5)$.
The model has a constant star formation rate and scale length (no inside out formation).
The black dots mark the evolution at age of 4, 8, 10, 11, 12 and 13 Gyr as in \autoref{fig:alpha_fe_Rz_one}.
\label{fig:alpha_fe_csfr}}
\end{figure}

\begin{figure*}[tb]
\centering \includegraphics[width=0.99\textwidth]{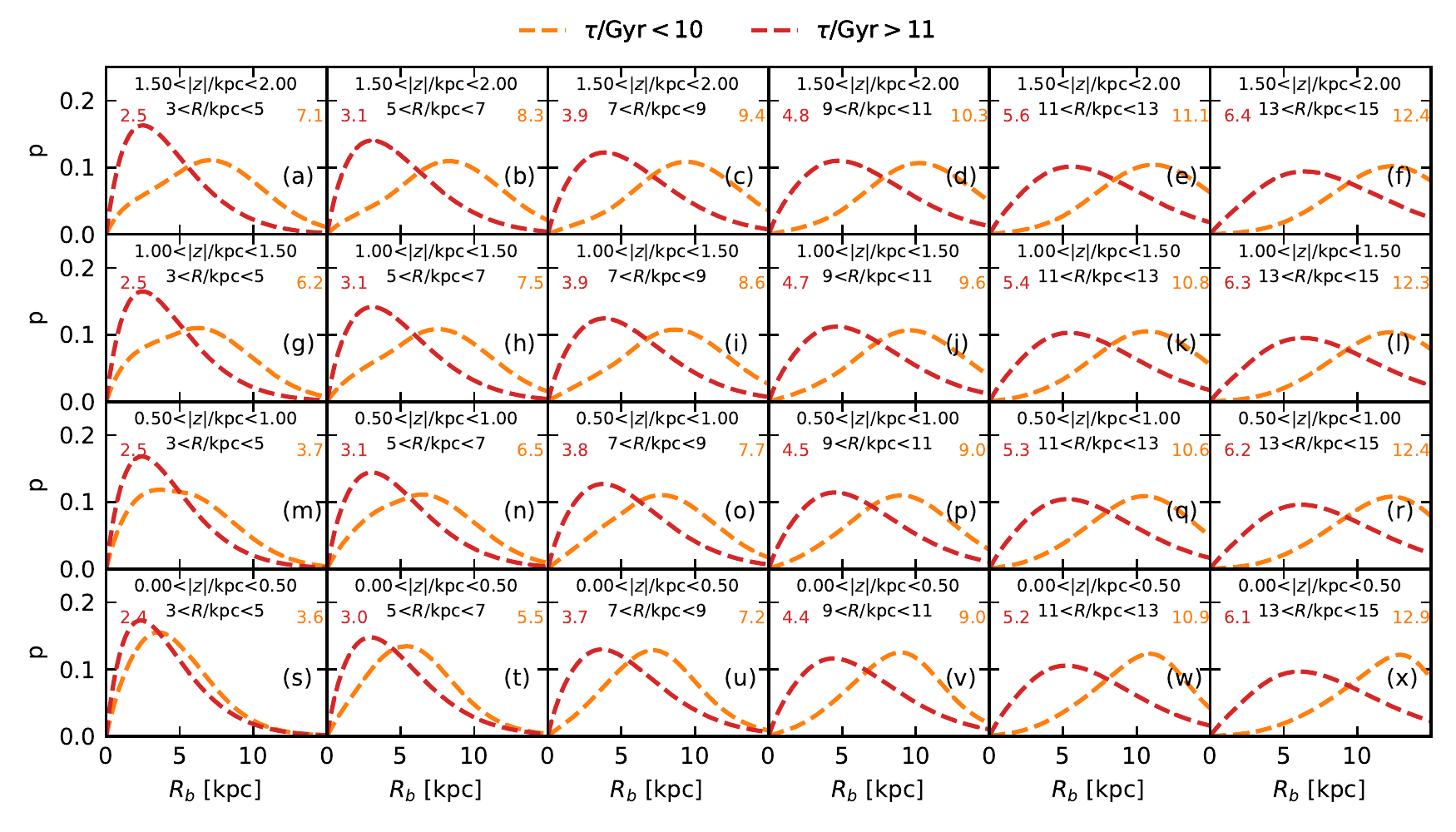}
\caption{Model predictions for the distribution of birth radius $R_b$. Shown are cases for young and old stars, which correspond to
low-\alphafe{} and high-\alphafe{} sequences respectively. The distributions correspond to star forming mass as selection function was not applied. The two numbers on each panel denote the position of the peaks in the distribution, left number for old stars and right number for young stars.
\label{fig:p_Rb_Rz}}
\end{figure*}

\section{Discussions}

\subsection{What is the reason for the existence of the high- and low-
\alphafe{} sequences?}

Comparing the map in \autoref{fig:alpha_fe_Rz_one} with the overlaid
evolutionary tracks provides insight into the origin of the high- and low-
\alphafe{} sequences. The high-\alphafe{} sequence coincides with the $R_b=4$ kpc
evolutionary track, suggesting that it is primarily a sequence of age.
This is very obvious at the high [Fe/H] end.
However, at the low [Fe/H] end, the high-\alphafe{} sequence is also partly a sequence of birth-radius.
In contrast, the low-\alphafe{} sequence follows a 8 Gyr isochrone, suggesting that
it is primarily a sequence of birth  radius.
The densest portions of both the sequences are parallel to the isochrones.
The dense low-\alphafe{} portion is made up of stars younger than 10 Gyr while
the dense high-\alphafe{} portion is made up of stars older than 11 Gyr.

The gap between the two sequences is due to the sharp transition of \alphafe{}
from a high value to a low value to, within a span of a few Gyrs and centered around 10.5 Gyr. This sharp transition, which is due to
time delay in the onset of SNIa explosions, creates a valley in the number density
of stars corresponding to the region occupied by the 10.5 Gyr isochrone, and
is the reason behind the existence of the double sequence.
In our preferred model the star formation history peaks at 10.5 Gyr and the radial scale length of the disc decreases with increasing age (\autoref{fig:amr_ism}).
In \autoref{fig:alpha_fe_csfr} we plot the \fehalpha{} distribution for a model
with constant star formation history and radial scale length, which looks
very similar to \autoref{fig:alpha_fe_Rz_one}. This demonstrates that
the double sequence is not due to any features in the profile of
star formation rate or scale length with age.

\subsection{How should we interpret the \fehalpha{} plane?} \label{sec:interpretation}
The best way to interpret the distribution of stars in the \fehalpha{} plane
for a given Galactic location is in terms of a sequence of evolutionary tracks corresponding to different birth radii (see \autoref{fig:alpha_fe_Rz_one}). For each point in \fehalpha{} plane there is a corresponding point in $(R_b,\tau)$ plane.
The tracks are labelled by their value for $R_b$ and $\tau$, which determines
the location of a point along the track. For a given Galactic location $(R,|z|)$,
the number density at an age $\tau$ on a track is given by
\be
p(R_b,\tau|R,z) &\propto& p(\tau)p(R_b|\tau)p(z|\sigma_z) \times \nonumber \\
&& [\int p(R_g|R_b,\tau)p(R|R_g,R_b,\tau) {\rm d}R_g]
\label{equ:ptrack}
\ee
The term in square bracket on the RHS is a function that typically peaks at around
$R_b=R$ but has tails extending to lower and higher $R_b$. The extent
of the tails is governed by the strength of churning ($\sigma_{L0}$) and blurring ($\sigma_R$).
In absence of churning and blurring, stars will be distributed only
along the track $R_b=R$ (similar to \autoref{fig:apogee_model_no_churn}).
The first term on the RHS of \autoref{equ:ptrack} is the star formation rate,
the second is the distribution of birth radii at the time of formation.
The third term is the vertical density, which is roughly proportional to ${\rm sech}^2(z/(2h_z))/h_z$, with scale height $h_z \propto \sigma_z^2$. $h_z$ in general increases with age (using \autoref{equ:f_tau} as $\tau^{0.88}$). At small $|z|$,
stellar populations with small $h_z$ will dominate, while at high $|z|$,
stellar populations with large $h_z$ will dominate. Going up the
evolutionary track, we expect to see kinematically hotter populations (populations with large $\sigma_z$). The rate of change of \alphafe{}
with age is highest at around 10.5 Gyr, which leads to
a local minimum in the number density of stars at that age.
Hence, even if $p(\tau)$ is smooth and continuous, we  will still see a minimum in the density distribution along a track and consequently a bimodality in the \fehalpha{} plane.
However, along a single evolutionary track,  $h_z \propto \tau^{0.88}$, the distribution
of $h_z$ is expected to be continuous as long as the star
formation rate is continuous.
It is now easy to see why the mass weighted distribution of scale height $h_z$
at the solar annulus can be a continuous function as found by \citet{2012ApJ...751..131B},
in spite of the \fehalpha{} distribution being bimodal.

\subsection{Why does the locus of the high-\alphafe{} sequence appear to be the same across all locations of the Galaxy?}
The high-\alphafe{} sequence seems to approximately follow the $R_b=4$ kpc evolutionary track
at all Galactic locations. It can be seen from \autoref{fig:alpha_fe_Rz_one}
that the shape of this track has a knee at ${\rm [Fe/H]} \sim -0.2$.
To the left of the knee the sequence is almost parallel to \feh{} axis and to the right
it slopes downwards. The left part is mainly made up of stars with age greater than 11 Gyr.
The lower envelope is made up of 11 Gyr isochrones with $R_b>4$ kpc which
runs parallel to the \feh{} axis. This is separated from the low-\alphafe{}
sequence due to the sharp transition of \alphafe{} when age changes from 11 to 10 Gyr. At different Galactic locations the distribution of $R_b$ can change but the high-\alphafe{}
sequence will still remain flat and the lower envelope will remain the same.
The upper/outer envelope is made up of stars with $R_b<4$ kpc. In this regime
the distribution of $R_b$ has a very characteristic shape, it is a rising function of $R_b$
at all locations. Hence, as long as there are enough stars coming from $R_b<4$ kpc,
we will always see the same shape of the outer envelope. The reason we always
have enough stars from the inner Galaxy is because of churning.
\autoref{fig:p_Rb_Rz} shows that the distribution of $R_b$ is a skewed distribution with a well defined peak. Such a distribution is predicted by \autoref{equ:rb_dist}, which peaks
at $R_b=R_d$. In reality, the peak shifts to larger $R_b$ with increase of $R$.
This is because the actual distribution of $R_b$ at a given $R$ (\autoref{equ:ptrack}) depends upon additional
factors involving churning and blurring.
For old stars ($\tau<10$ Gyr) the churning is very efficient such that
the term within the square bracket in \autoref{equ:ptrack} is a very broad, hence the $p(R_b|\tau)$ term dominates.

\citet{2014ApJ...796...38N} argue that thinness of the high-\alphafe{} sequence combined with the fact that the same
sequence exists at a wide range of $R$ and $|z|$,
is probably indicative of the fact that similar condition existed throughout the disc. This argument
was further supported by \citet{2016ApJ...823...30B} based on similarity of radial profile of high-\alphafe{} MAPs.
However, the existence of the high-\alphafe{} sequence at all $R$ and $|z|$ says very little about where they were born. Birth radius is uniquely specified by location on the \fehalpha{} plane.
The fact that the high-\alphafe{} sequence can still be seen at large $R$ is due to churning and blurring. For ages greater than 11 Gyr, different evolutionary tracks have very similar \alphafe{} at birth. This is the reason for the thinness along the high-\alphafe{} track: a spread in birth radius does not have a spread in \alphafe{}, only \feh. The high-\feh{} end of the sequence is made up of stars with $R_b<4$ kpc, here the evolutionary tracks are close together in the \fehalpha{} plane and appear quite similar.

\subsection{Chemical enrichment}
We have proposed an empirical model for chemical enrichment and constrained
it using observational data containing age and abundance of stars. The fact that it
reproduces the distribution of stars in \fehalpha{} plane at different Galactic
locations, further supports the adopted enrichment model. However,
our model is also physically motivated and its parameters can be used to
shed light on the physics of enrichment. $\Delta t_{\alpha}$ indicates
the time delay between the onset of star formation and the peak in the rate of
SNIa, which mainly depends on the lifetime of the binary companion of a white dwarf.
Our adopted value of 1.5 Gyr is in good agreement with typical
expectation from theoretical models \citep{2003MNRAS.340..908K}. Our results suggest
that the initial composition at all birth radius was very similar,
probably due to short timescale of SNII that sets the initial value
of \alphafe{}. At later times,  the \feh{} and \alphafe{} values are found
to reach an equilibrium value which depends on birth radius, this is
probably regulated by gas dynamical processes like inflow of fresh gas,
outflows and radial flows, but needs further investigation.
In the future, we should relax some of the assumptions that were made
and let the data inform if they are true.

\subsection{The role played by velocity dispersion relations}
The overall pattern of the \fehalpha{} distributions is sensitive to the velocity dispersion relations,
specially, the relative fraction of stars in the high- and the low-\alphafe{} sequences.
This is because $\sigma_z$ determines the scale height of a
population and the scale height determines which population is going
to dominate at what $|z|$ (\autoref{sec:interpretation}).
The high-\alphafe{} sequence
is primarily made up of old stars while the low-\alphafe{} sequence
is made up of comparatively younger stars. So their relative fraction is particularly sensitive to the dependence of $\sigma_z$ on age.

For a given $R$, with increase of $|z|$, the peak for the low-\alphafe{} sequence moves to lower [Fe/H] and higher \alphafe{}, both for the observations and the model. This is most evident for panels corresponding to $7<R/{\rm kpc}<9$ (\autoref{fig:apogee_data} and \autoref{fig:apogee_model}).
The slight increase in \alphafe{} of the peak is due to increase of $\sigma_z$ with
age, which makes it more likely for older stars to occupy regions with higher $|z|$.
But the shift of \feh{} is more than that predicted by simply travelling up along the
orange evolutionary track.
The extra shift of the peak to lower [Fe/H] is specifically due to anti-correlation of
$\sigma_z$ with [Fe/H], which makes it more likely for low [Fe/H] stars
to occupy regions with higher $|z|$. If the parameter $\gamma_{{\rm [Fe/H]},v}$
that controls the dependence of $\sigma_z$ on [Fe/H] is set to zero, no shift
of the peak is seen in the distributions predicted by the model.
This provides an independent confirmation of the \citet{2020arXiv200406556S} scaling of velocity dispersion with \feh{} (or equivalently birth radius) for a given age.

In the topmost $|z|$ slice (\autoref{fig:apogee_data} and \autoref{fig:apogee_model}), we have a dominant low-\alphafe{} sequence at large $R$ .
Naively, we expect the topmost slice to be dominated by old stars as they have high $\sigma_z$
and hence large scale height. We see the domination of old, high-\alphafe{} stars for $R<9$ kpc but not for larger values of $R$.
Inside out formation of the disc, i.e., scale length of a disc
being smaller at earlier times, is one possible explanation for this result.
In our model, $R_d^{\rm min}<R_d^{\rm max}$ indicates
inside out formation of the disc.
However, setting the $R_d^{\rm min}=R_d^{\rm max}$ was found to have very little effect
on the distributions.
The high-\alphafe{} sequence was found to shift slightly towards
low [Fe/H], which is due to more contribution from stars with larger $R_b$,
but for $R>9$ kpc the low-\alphafe{} sequence was still dominant.
Next, we set the parameter $\alpha_{L,v}$, which regulates the increase of $\sigma_z$ with $L$, to zero. With this change, the high-\alphafe{} sequence
was found to dominate the panels (d), (e), and (f) corresponding to large
$R$. $\alpha_{L,v}$ is responsible for
flaring of young low-\alphafe{} stars in the outer disc, and this makes
the low-\alphafe{} sequence dominate at large $R$.
Low-\alphafe{} stars are made up of all stars with age less than 10 Gyr,
hence, they significantly outnumber the high-\alphafe{} stars. So, even a small
amount of flaring is enough to make them dominate over the high-\alphafe{} stars.

Flaring has been reported in the outer disc of the Milky Way and has been shown to
occur in numerical simulations \citep{2015ApJ...804L...9M}.
Both \citet{2016ApJ...823...30B} and \citet{2017MNRAS.471.3057M} using APOGEE data showed that the youngest stars flare the most. Simulations of discs with GMCs, spiral arms and a bar by \citet{2017MNRAS.470.2113A} show that flaring is due to constant
birth velocity dispersion.
\citet{2020arXiv200406556S} provide kinematic indication of flaring.
For constant scale height, $\sigma_z$ is expected to decrease exponentially with $L$.
\citet{2020arXiv200406556S} show that $\sigma_z$
decreases with $L$ for upto solar angular momentum, but increases thereafter, which is
indicative of flaring in the outer disc. They argue that constant birth dispersion
will lead to more flaring in young stars and flaring will start at much lower values of $R$.  The reason being that flaring starts when an exponentially
declining $\sigma_z$ as a function of $L$ hits the floor of constant birth dispersion, for younger stars the overall dispersion is small and hence the floor is hit at a smaller $L$.

\subsection{Relation to other studies}
\citet{2009MNRAS.396..203S} introduced a detailed model for chemical evolution
with radial migration and gas flows that was capable of simulating
the joint distribution of abundances and phase space coordinates.
They made predictions for the distribution of solar neighborhood stars
in the ${\rm([Fe/H], [O/Fe])}$ plane. They made predictions for main sequence
type stars following the GCS survey \citep{2004A&A...418..989N} selection function.
Note, for the purpose of the
discussion here, ${\rm [O/Fe]}$ can be considered as proxy for \alphafe{}.
Their models were able to reproduce the high- and
low-\alphafe{} sequences. They showed that the double sequence
has nothing to do with breaks in star formation history but was
a consequence of the sharp enrichment of \alphafe{} due to SNIa.
The low-\alphafe{} sequence was a sequence of stars born at different
radius but were present in the solar neighborhood due to radial migration.
They also predicted an anti-correlation of angular momentum
with \feh{} which was in qualitative agreement with data from GCS.
We arrive at the same conclusions, but using an empirical model
for chemical enrichment that is calibrated to observational data
and using an improved model for velocity dispersion.
Our empirical relations for the evolution of abundances are in good
agreement with theirs, which provides strong support to their
ab-inito chemical evolution model.
They compared their results with a small observational data set,
moreover, the observational data set had kinematic biases.
Hence it was not possible to do a proper comparison of density distribution in the \alphafe{} plane
with the models. Specifically, the models predicted a significant number of
stars in between the two sequences but that seemed to be missing in the
presented observational data.
In contrast, we make a detailed comparison with observations using
a significantly larger data set and test the predictions
over different locations across the Galaxy.

\citet{2013A&A...549A.147B,2017A&A...605A..89B} studied the abundance
distribution of bulge stars using microlensed dwarfs and subgiants
within 1 kpc of the Galactic center. They found that \fehalpha{} distribution of stars in the Galactic bulge are very similar to the sequence found in the inner disc \citep[see also][]{2019A&A...626A..16R}
. This is correctly predicted by our model. We expect the bulge to look like panel (m) and (s) of \autoref{fig:apogee_model}.The bulge should follow the $R_b<1$ kpc evolutionary track, and this is similar to $R_b=4$ kpc track which represents the high-\alphafe{} sequence seen throughout the Galaxy.

\citet{2016ApJ...823...30B} studied the $R$ and $|z|$ distributions
of mono-abundance populations using APOGGE red-clump
stars. They found  that the radial distribution of high-\alphafe{} MAPs
are well described by a single exponential
(for $R>4$ kpc) but that of low-\alphafe{} MAPs
are more complex. The low-\alphafe{} MAP stars are distributed in a ring like
structure characterized by a peak with exponential fall off away from the
peak both for smaller and larger radii.
\citet{2017MNRAS.471.3057M}, using mono-age and mono-metallicity APOGEE red-giant-branch stars, also reported similar findings.
The above findings are easy to understand
using \autoref{fig:alpha_fe_Rz_one}. A low-\alphafe{} MAP represents
young stars born at a particular birth radius. Our churning mechanism
predicts that at any given time after birth, the distribution of guiding radius
$p(R_g|R_b,\tau)$ will be ring-like centered around $R_b$.
Blurring further distributes stars with a given $R_g$ over $R$ in a ring around $R_g$.
Hence, the distribution of $R$,
\be
p(R|R_b,\tau)= \int p(R_g|R_b,\tau) p(R|R_g,\tau,R_b) {\rm d}R_g
\ee
will
also be ring-like as it is given by a convolution of one ring-like
distribution with another ring-like distribution.
A high-\alphafe{} MAP typically represents stars with $R_b$ of about
4 kpc, which will also be ring like but with peak close to $R=4$ kpc.
Since there was no observed data inwards of 4 kpc, the
radial distribution was expected to be well fit by a single exponential.
Along the high-\alphafe{} sequence, the evolutionary tracks
are closely spaced. Hence, a MAP can in general also contain stars from multiple birth radii, which can shift the peak further inwards.

\citet{2015MNRAS.449.3479S} proposed action-based analytical
distribution function with a prescription for radial migration.
Our model is similar to theirs, but unlike them
we also make predictions for \alphafe{}.
We adopt their prescription for radial migration. The strength of
migration is characterized by parameter $\sigma_{L0}$, defined as the
dispersion of angular momentum for 12 Gyr population.
By fitting to GCS stars, \citet{2015MNRAS.449.3479S}
estimated $\sigma_{L0}=1150$ kpc km/s.
We adopt a very similar value and find that it reproduces the
\fehalpha{} distribution of APOGEE stars quite well.
Recently, \citet{2020arXiv200204622F} have also estimated
$\sigma_{L0}$ making use of APOGEE red-clump stars and
building upon the model of \citet{2015MNRAS.449.3479S}.
They estimate the dispersion for a 12 Gyr population to be 875 kpc km/s,
which is slightly smaller.

\section{Conclusions}
We have presented an analytical chemodynamical model of the Milky Way that can make predictions for the joint distribution of position, velocity, age and abundance of stars in the Milky Way.
Parametric models of this sort have important uses. Even before the Gaia DR2 data release, observational multi-dimensional data sets were becoming vast and unwieldy. The same holds true for cosmological simulations of Milky Way analogues \citep[e.g. ][]{2018MNRAS.480..652E}. Our model provides a framework for fitting both observational and simulated data, and tying both together through a basis set of key parameters.

The key aspect in which the model improves upon previous
works is its inclusion of a new prescription for the evolution
of \alphafe{} with age and \feh{} and a new
set of relations describing the velocity dispersion of stars.
For the first time, we have been able to show that a model
with a smooth and continuous star formation history and velocity dispersion relations
can reproduce
the \fehalpha{} distribution of observed stars at different $R$
and $|z|$ locations across the Galaxy.
The model also satisfies a number of other observational constraints.
It has a vertical distribution
of stars that is well fit by a sum of two exponential functions.
For the low-\alphafe{} stars, the model is also able to reproduce the
trend of mean angular momentum as a function of metallicity at different $R$
and $|z|$ locations.

A number of finer details of the observed \fehalpha{} distribution  are also correctly reproduced. These include (i) the observed double sequence (bimodality)
in the \fehalpha{} plane, (ii) the relative fraction of stars in the
high- and the low-\alphafe{} sequences and its variation with $R$ and $|z|$,  (iii) the change in position of low-\alphafe{} peak with $R$ and $|z|$, and (iv) the skewness
of the MDFs as a function of $R$ and $|z|$.
Our work confirms and significantly extends the earlier findings of
\citet{2009MNRAS.396..203S} relating to the origin of the double sequence in \fehalpha{} plane and the thick disc; their study was limited to the
solar neighborhood and lacked a detailed comparison with greatly improved data since the Gaia DR2 data release. Our work is also in agreement with
\citet{2012ApJ...751..131B} who had shown that the scale-height distribution of mono-abundance populations is continuous, which supports the argument that the star formation history is also continuous.
The ring-like radial distribution of stars for a low-\alphafe{} mono-abundance populations
as shown by \citet{2016ApJ...823...30B} and \citet{2017MNRAS.471.3057M}
is also in agreement with the predictions of our model.

In \citet{2020arXiv200406556S}, it was shown that for older stars the apparent break and rise of the velocity dispersion profile, with respect to that
of a power law, is due to systematic decrease of angular momentum with radius.
When this is taken into account, the velocity dispersion
of old and high-\alphafe{} stars, which are traditionally associated with the thick disc, also
follow the same set of relations for their dependence on age, angular momentum and
metallicity as that of other stars that make up the thin disc.
Hence, the break in age velocity-dispersion relation, the bimodality in \fehalpha{} distribution, the uniformity of the locus of the high-\alphafe{} sequence, and the double exponential nature of the vertical density distribution, do not require an abrupt change in either the star formation history or the kinematic evolutionary history of the Milky Way. In other words, these are no longer
sufficient arguments for the existence of a distinct thick disc stellar population. The word `distinct' is used in the sense that the evolution is not smooth or continuous.
A brief period of quenching, as proposed by others \citep{2019A&A...625A.105H,2001ApJ...554.1044C}, could potentially be still present
but it is not required to explain the above mentioned properties of the Milky Way.

The high-\alphafe{} sequence at the low-\feh{} end is a sequence of both age and birth radius, while at the high-\feh{} end it is a sequence of age.
In contrast, the low-\alphafe{} sequence is primarily a trend of different birth radius.
The origin of the double sequence is due to two key processes: the sharp transition the
of \alphafe{} at around 10.5 Gyr ago, and the radial migration of stars.
The transition is most likely due to the delay between the onset of star formation and the occurrence of SNIa in the early Universe. This sharp transition creates a valley in the density distribution of stars in the \fehalpha{} plane, approximately parallel to the \feh{} axis.
The radial migration, more precisely the process of churning, is responsible for
the large spread of the low-\alphafe{} sequence along the \feh{} axis.
We show that if churning is not included the process of blurring alone is not sufficient to reproduce the double sequence. At any given radius,
the high-\alphafe{} sequence is dominated by stars that have migrated outwards from the
inner Galaxy, however, the contribution of locally born stars and inward migrators is not negligible.

The apparent uniformity in the locus of the high-\alphafe{} sequence is due
to churning being very efficient. Efficient churning firstly makes it possible
for enough stars from the inner radius to reach large $R$,
and secondly it makes the distribution of birth radius $R_b$ almost independent of $R$.

The velocity dispersion relations are responsible for some of the
systematic trends of the \fehalpha\ distribution with $R$ and $|z|$. The MDF of the low-\alphafe{} sequence is found to shift towards lower \feh{}
with increase of $|z|$. This is due to dependence of vertical velocity
dispersion on \feh{} (or equivalently birth radius). At high $|z|$, the
lack of high-\alphafe{} stars for $R>9$ kpc, is due to flaring, and the flaring is due to the
parabolic shaped dependence of velocity dispersion on
angular momentum, characterized by a minimum at around
solar angular momentum and a rise thereafter.

There are various aspects of the model that can be improved
in the future. We have only explored the evolution
of iron and $\alpha$ elements, it should be straightforward
to extend the model to also include $r$ and $s-$process elements.
Production sites and nucleosynthetic yields for these elements
are not well understood. High resolution spectroscopic surveys
like GALAH and APOGEE are now providing reliable estimates of
abundances for these elements for a large number of stars
in the Milky Way. The phase-space evolution of the model
is now reasonably well constrained, in future, we can focus on the physics exclusive to these elements.

Our model has a number of free parameters
and there are likely to be degeneracies between some of them,
which we have not explored. The parameters were tuned manually.
In the future, a proper MCMC based exploration of the parameter space should be useful \citep[e.g. ][]{2017ARA&A..55..213S}.

One of the greatest strengths of the model is that,
being purely analytical, it can be easily fit to observational or cosmologically simulated data. Since the model is physically motivated, it means that we can
gain understanding about the various physical processes that have
shaped our Galaxy. One of the most poorly
understood physical processes is radial migration. For our
parametric model, we have adopted only a simple prescription.
In reality, a more complex process is likely to exist as non-axisymmetric perturbations (spiral arms, bars, interlopers) come and go over the aeons.

Our results show that the distribution of low-\alphafe{} stars in the \fehalpha\ plane and their variation with $R$ and $|z|$, is very
sensitive to radial migration, this is very promising to
constrain radial migration.
The gap between the low-\alphafe{} and the high-\alphafe{} sequence
is very sensitive to parameters $t_{\rm \alpha}$ and $\Delta t_{\alpha}$
that control the enrichment of $\alpha$ elements in the Galaxy,
another process that is not fully understood.

Our chemical evolution model, although physically motivated, is still empirical in nature. The star formation rate is decoupled from the chemical evolution which is clearly incorrect.
In future, it will be useful to investigate {\it ab initio} chemical evolution models, that take star formation, gas infall, outflows and nucleosynthesis and fine-tune them to reproduce the chemical evolution tracks that we have derived here.
Finally, the model being purely analytical, it should be easy to
insert into forward-modelling tools like {\sl Galaxia} \citep{2011ApJ...730....3S} that generate synthetic catalogs of stars and are useful for interpreting stellar surveys.

\acknowledgments
SS is funded by a Senior Fellowship (University of Sydney), an ASTRO-3D Research Fellowship and JBH's Laureate Fellowship from the Australian Research Council (ARC).
JBH's research team is supported by an ARC Laureate Fellowship (FL140100278) and funds from ASTRO-3D. MJH is supported by an ASTRO-3D 4-year Research Fellowship.

The GALAH Survey is supported by the ARC Centre of Excellence for All Sky Astrophysics in 3 Dimensions (ASTRO 3D), through project CE170100013.
This work has made use of data acquired through the Australian Astronomical Observatory, under programs: GALAH, TESS-HERMES and K2-HERMES. We acknowledge the traditional owners of the land on which the AAT stands, the Gamilaraay people, and pay our respects to elders past and present.

This work has made use of data from the European Space Agency (ESA) mission
{\it Gaia} (\url{https://www.cosmos.esa.int/gaia}), processed by the {\it Gaia}
Data Processing and Analysis Consortium (DPAC,
\url{https://www.cosmos.esa.int/web/gaia/dpac/consortium}). Funding for the DPAC
has been provided by national institutions, in particular the institutions
participating in the {\it Gaia} Multilateral Agreement.

This work has made use of data from SDSS-III.
Funding for SDSS-III has been provided by the Alfred P. Sloan Foundation, the Participating Institutions, the National Science Foundation, and the U.S. Department of Energy Office of Science. The SDSS-III web site is http://www.sdss3.org/.

This work has made use of Guoshoujing Telescope (the Large Sky Area Multi-Object Fiber Spectroscopic Telescope LAMOST) which is a National Major Scientific Project built by the Chinese Academy of Sciences. Funding for the project has been provided by the National Development and Reform Commission. LAMOST is operated and managed by the National Astronomical Observatories, Chinese Academy of Sciences.

\bibliographystyle{yahapj}
\bibliography{references}
\end{document}